\documentclass[pdflatex,sn-mathphys-num]{sn-jnl}

\usepackage{graphicx}%
\usepackage{multirow}%
\usepackage{amsmath,amssymb,amsfonts}%
\usepackage{amsthm}%
\usepackage[scaled=1.0]{rsfso}
\usepackage[title]{appendix}%
\usepackage{xcolor}%
\usepackage{textcomp}%
\usepackage{manyfoot}%
\usepackage{booktabs}%
\usepackage{algorithm}%
\usepackage{algorithmicx}%
\usepackage{algpseudocode}%
\usepackage{listings}%
\usepackage{xspace}
\usepackage{url}
\usepackage{hyperref}

\usepackage{xr}
\makeatletter
\newcommand*{\addFileDependency}[1]{
  \typeout{(#1)}
  \@addtofilelist{#1}
  \IfFileExists{#1}{}{\typeout{No file #1.}}
}
\makeatother

\theoremstyle{thmstyleone}%
\theoremstyle{thmstyletwo}%

\theoremstyle{thmstylethree}%

\newcommand{\LCO}{LiCoO\textsubscript{2}\xspace}
\newcommand{\LqCO}{Li\textsubscript{1/3}CoO\textsubscript{2}\xspace}
\newcommand{\LxCO}{Li\textsubscript{x}CoO\textsubscript{2}\xspace}

\newcommand{\CoII}{Co\textsuperscript{2+}\xspace}
\newcommand{\CoIII}{Co\textsuperscript{3+}\xspace}
\newcommand{\CoIV}{Co\textsuperscript{4+}\xspace}

\newcommand{\LNO}{LiNiO\textsubscript{2}\xspace}
\newcommand{\LxNO}{Li\textsubscript{x}NiO\textsubscript{2}\xspace}
\newcommand{\NO}{NiO\textsubscript{2}\xspace}
\newcommand{\NdNO}{NdNiO\textsubscript{3}\xspace}
\newcommand{\Nioct}{NiO\textsubscript{6}\xspace}

\newcommand{\NiIII}{Ni\textsuperscript{3+}\xspace}
\newcommand{\NiIV}{Ni\textsuperscript{4+}\xspace}

\newcommand{\LFPO}{LiFePO\textsubscript{4}\xspace}

\newcommand{\FeII}{Fe\textsuperscript{2+}\xspace}
\newcommand{\FeIII}{Fe\textsuperscript{3+}\xspace}

\newcommand{\dsix}{\textit{d}\textsuperscript{6}\xspace}

\newcommand{\deight}{\textit{d}\textsuperscript{8}\xspace}
\newcommand{\dsixL}{\textit{d}\textsuperscript{6}\underline{\textit{L}}\xspace}
\newcommand{\dsevenL}{\textit{d}\textsuperscript{7}\underline{\textit{L}}\xspace}
\newcommand{\deightL}{\textit{d}\textsuperscript{8}\underline{\textit{L}}\xspace}

\newcommand{\dsevenLt}{\textit{d}\textsuperscript{7}\underline{\textit{L}}\textsuperscript{2}\xspace}
\newcommand{\deightLt}{\textit{d}\textsuperscript{8}\underline{\textit{L}}\textsuperscript{2}\xspace}
\newcommand{\deightLthree}{\textit{d}\textsuperscript{8}\underline{\textit{L}}\textsuperscript{3}\xspace}
\newcommand{\Ledge}{\textit{\textit{L}\textsubscript{3}}$-$edge\xspace}
\newcommand{\Kedge}{\textit{K}$-$edge\xspace}
\newcommand{\Ogas}{O\textsubscript{2}\xspace}
\newcommand{\eg}{e\textsubscript{g}\xspace}

\begin{document}

\title[Negative Charge Transfer: Ground State Precursor towards High Energy Batteries ]{Negative Charge Transfer: Ground State Precursor towards High Energy Batteries }


\author[1,2]{\fnm{Eder~G.} \sur{Lomeli}}
\equalcont{These authors contributed equally to this work.}

\author[3,4,5]{\fnm{Qinghao} \sur{Li}}
\equalcont{These authors contributed equally to this work.}

\author[1,2]{\fnm{Kuan~H.} \sur{Hsu}}
\equalcont{These authors contributed equally to this work.}

\author[3]{\fnm{Gi-Hyeok} \sur{Lee}}
\author[3]{\fnm{Zengqing} \sur{Zhuo}}
\author[6]{\fnm{Bryant-J.} \sur{Polzin}}
\author[6]{\fnm{Jihyeon} \sur{Gim}}
\author[6]{\fnm{Boyu} \sur{Shi}}
\author[6]{\fnm{Eungje} \sur{Lee}}
\author[7]{\fnm{Yujia} \sur{Wang}}
\author[7]{\fnm{Haobo} \sur{Li}}
\author[7]{\fnm{Pu} \sur{Yu}}
\author[2,3]{\fnm{Jinpeng} \sur{Wu}}
\author[2,8,9]{\fnm{Zhi-Xun} \sur{Shen}}
\author[4]{\fnm{Shishen} \sur{Yan}}
\author[1]{\fnm{Lauren} \sur{Illa}}
\author[2,10]{\fnm{Josh~J.} \sur{Kas}}
\author[2,10]{\fnm{John~J.} \sur{Rehr}}
\author[11]{\fnm{John} \sur{Vinson}}
\author[2]{\fnm{Brian} \sur{Moritz}}
\author[3]{\fnm{Yi-Sheng} \sur{Liu}}
\author[3]{\fnm{Jinghua} \sur{Guo}}
\author[3]{\fnm{Yi-de} \sur{Chuang}}

\author*[3]{\fnm{Wanli} \sur{Yang}}\email{wlyang@lbl.gov}
\author*[1,2,8]{\fnm{Thomas P.} \sur{Devereaux}}\email{tpd@stanford.edu}

\affil[1]{\orgdiv{Department of Materials Science and Engineering}, \orgname{Stanford University}, \orgaddress{ \city{Stanford}, \state{CA}, \country{USA}}}

\affil[2]{\orgdiv{Stanford Institute for Materials and Energy Sciences}, \orgname{SLAC National Accelerator Laboratory}, \orgaddress{\city{Menlo Park}, \state{CA}, \country{USA}}}

\affil[3]{\orgdiv{Advanced Light Source}, \orgname{Lawrence Berkeley National Laborator}, \orgaddress{\city{Berkeley}, \state{CA}, \country{USA}}}

\affil[4]{\orgdiv{School of Physics, National Key Laboratory of Crystal Materials,}, \orgname{Shandong University}, \orgaddress{\city{Jinan}, \country{China}}}

\affil[5]{\orgdiv{College of Physics}, \orgname{Qingdao University}, \orgaddress{\city{Qingdao}, \country{China}}}

\affil[6]{\orgdiv{Chemical Sciences and Engineering Division}, \orgname{Argonne National Laboratory}, \orgaddress{\city{Lemont}, \state{IL}, \country{USA}}}

\affil[7]{\orgdiv{State Key Laboratory of Low Dimensional Quantum Physics and Department of Physics}, \orgname{Tsinghua University}, \orgaddress{\city{Beijing}, \country{China}}}

\affil[8]{\orgdiv{Geballe Laboratory for Advanced Materials}, \orgname{Stanford University}, \orgaddress{\city{Stanford}, \state{CA}, \country{USA}}}

\affil[9]{\orgdiv{Department of Physics and Applied Physics}, \orgname{Stanford University}, \orgaddress{\city{Stanford}, \state{CA}, \country{USA}}}

\affil[10]{\orgdiv{Department of Physics and Applied Physics}, \orgname{University of Washington}, \orgaddress{\city{Seattle}, \state{WA}, \country{USA}}}

\affil[11]{\orgdiv{Material Measurement Laboratory}, \orgname{National Institute of Standards and Technology}, \orgaddress{\city{Gaithersburg}, \state{MD}, \country{USA}}}


\abstract{Modern energy applications, especially electric vehicles, demand high energy batteries. However, despite decades of intensive efforts, the highest energy density and commercially viable batteries are still based on \LCO, the very first generation of cathode materials. The technical bottleneck is the stability of oxide-based cathodes at high operating voltages. The fundamental puzzle is that we actually never understood the redox mechanism of \LCO. Conventional wisdom generally defines redox to be centered on cations at low voltages~\cite{goodenough_review}, and on anions, \textit{i.e.} oxygen, at high voltages by forming oxidized chemical states like O$_2$ or peroxo-species~\cite{Sathiya2013,Grimaud2016,house_superstruct,House_o2,House_delocalO2,House_trappedo2,Tarascon_Peroxo,Seo2016}. Here, through \textit{in-situ} and \textit{ex-situ} spectroscopy coupled with theoretical calculations, we show that high-energy layered cathodes, represented by \LCO and \LNO, operate through enhancement of negative charge transfer (NCT) ground states upon charging throughout the whole voltage range. i.e., NCT evolution itself is the intrinsic redox mechanism regardless of voltage ranges.  NCT inherently engages high covalency and oxygen holes, leading to optimized performance without conventional redox centers in \LCO. The level of NCT, i.e., number of ligand holes, naturally explains many seemingly controversial results. The redefinition of redox mechanism reveals the pathway toward viable high energy battery electrodes.}




\maketitle
\section{Introduction}\label{sec:intro}

\noindent Modern energy applications like electric vehicles and smart grids have triggered a pressing demand for high energy-density batteries with improved capacity and power (voltage). However, achieving a cyclable battery with high operating voltage remains a formidable challenge. While many limiting factors involve the electrolyte and anode, a critical issue is the stability of transition-metal (TM) oxide-based battery cathodes at high voltages. The original invention of oxide-based cathodes, rather than sulfides, was triggered by studies of \LNO, which leads to the very first generation of commercial cathodes based on \LCO in 1991. The fundamental concept behind this genius is to maintain the redox center on the TM cations through “redox pinning”~\cite{goodenough_review}. Over the last decade, however, intensive efforts have been reported on utilizing the oxygen anions as the redox center to harvest more capacity at higher voltages~\cite{Sathiya2013,Grimaud2016}. Anionic redox is widely believed to originate from the formation and decomposition of chemically defined oxidized oxygen states through dimerization~\cite{house_superstruct,House_o2,House_delocalO2,House_trappedo2,Tarascon_Peroxo,Seo2016}, which often leads to detrimental \Ogas and/or radical oxygen release. Other than the focus on such unhybridized chemical forms, a few recent studies have emphasized the important role that TM-O hybridization plays in anionic redox~\cite{Kitchaev_pi_redox,Grey_LNO_Joule,ogley_metal_ligand,degroot_lco,Tsuchimoto_NMO,yang_cation_matter,Barbiellini_compton},  but the behavior at high voltages remains a hotly debated topic with controversial reports. Nevertheless, the battery community is dominated by a baseline mechanistic understanding of high-voltage cathodes: TM redox dominates the reversible low-voltage range, while reversible anionic redox is possible at higher voltages, but often has issues with stability and transport kinetics~\cite{Tarascon_NE_review}.

In contrast to the clash of opinions over anionic redox at high voltages, there has been common acceptance on TM redox at low voltages. However, even 45 years after its invention, experiments have never clarified such a mechanism for \LCO. Electron paramagnetic resonance (EPR) measurements suggested there was no Co redox at all in \LCO~\cite{Niemoller_lco_epr}, interpreted as an indication of ``pure oxygen redox''. Element sensitive spectroscopies have never revealed clear signatures of Co redox in \LCO; rather, spectroscopic results show only changes in features associated with Co-O hybridized states throughout the whole voltage range~\cite{degroot_lco,Mizokawa_LCO_XAS,Ensling_lco_band}. Recent theoretical analysis also challenges the assignment of Ni spectral changes as evidence of Ni redox in \LNO~\cite{Grey_LNO_Joule}, calling into question the conventional unified picture of redox in these layered oxide cathodes, which still remain the most promising candidates for high-energy batteries.

In this work, we targeted a holistic model of both the high- and low-voltage redox behavior of layered oxide-based high-energy cathodes. We revisited the redox mechanism in \LCO using valence-sensitive Co \Ledge and O \Kedge spectroscopic probes with in-depth theoretical analysis of both charged and discharged states. We developed for the first time soft X-ray \textit{in-situ}/\textit{operando} \LCO/Li cells with commercial electrolyte to rule out technical artifacts in soft X-ray absorption spectroscopy (sXAS). Numerical modeling, compared with results of sXAS and resonant inelastic X-ray scattering (RIXS), allowed us to achieve a quantitative understanding of the electrochemical evolution of Co and O states. Throughout the voltage range from 3.0~V to 4.8~V, we found that the charge compensated ground state wavefunction is consistent with an enhanced negative charge transfer (NCT), where oxygen is partially oxidized and the 2p shell is not full.
NCT also inherently leads to 
unusually strong hybridization, consistent with spectroscopic observations of highly itinerant states upon charging. Although NCT signatures in our RIXS results are unambiguous and the majority of the electron loss occurs on oxygen, dimerization features could only be detected once the voltage increases above the level where a notorious structural phase transition takes place in NCT model compounds. We performed similar analysis on \LNO to provide more universality to our observation of NCT ground states and fundamentally reshape our understanding of the redox mechanism in layered oxide cathodes.

\section{Results}

\subsection{\textit{Operando} and \textit{ex-situ} sXAS of \texorpdfstring{LiCoO$_2$}{LiCoO2}}

\begin{figure} [!hbt]
	\centering
	\includegraphics[width=0.9\textwidth]{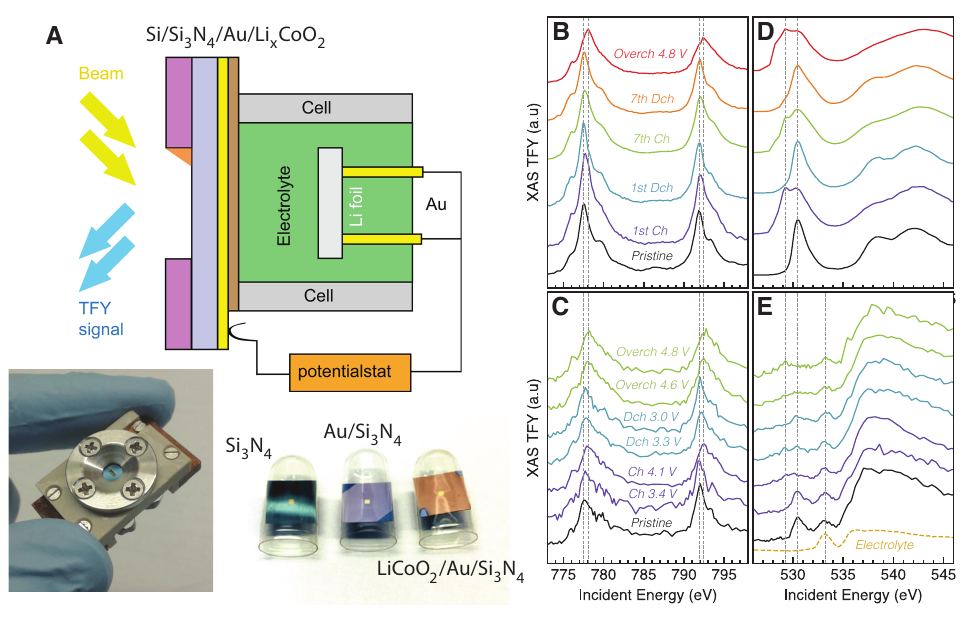} 
	\caption{\textbf{\textit{Operando} vs \textit{ex-situ} sXAS of cycled \LxCO.}
		(\textbf{A}) Schematic and photos of soft X-ray \textit{in-situ} cells used for \textit{operando} experiments. The \LCO thin film deposited on 100~nm $\mathrm{Si_3N_4}$/Au membrane was utilized as the cathode and Li foil as counter and reference electrode. Commercial $\mathrm{LiPF_6}$/EC:EMC:DEC (4:2:4) is used as the liquid electrolyte. (\textbf{B}) Typical \textit{ex-situ} Co $L-$edge sXAS TFY spectra collected at representative electrochemical states: 1st charged to 4.2~V, 1st discharged to 3.0~V, 7th charged to 4.2~V, 7th discharged to 3.0~V and overcharged to 4.8~V. (\textbf{C}) \textit{Operando} Co $L-$edge sXAS spectra collected at representative electrochemical states with cycling voltages as noted. (\textbf{D}) Typical \textit{ex-situ} O \Kedge sXAS TFY spectra. (\textbf{E}) \textit{Operando} O \Kedge sXAS collected at representative electrochemical potentials as indicated. Dashed lines on (\textbf{B-E}) are guides to the eye indicating the main features.}
	\label{fig:LCO_exp} 
\end{figure}

Soft X-ray Co $L$-edges correspond to excitations directly into the TM 3$d$ valence states~\cite{qinghao_xray_redox}, but measurements have not been performed on real-world electrodes in \textit{operando}, thus calling into question whether the missing signatures of Co redox are intrinsic. We thus fabricated an \textit{operando} electrochemical cell (Fig.~\ref{fig:LCO_exp}A) with \LCO thin films by optimizing the sealing mechanism for corrosive commercial electrolyte (LiPF\textsubscript{6}/EC:EMC:DEC 4:2:4) based on previously developed soft X-ray \textit{in-situ} cells~\cite{jinghua_RIXS_review}. The technical details and electrochemical profiles for the \LCO thin film are shown in the methods section and Fig.~\ref{fig:lco_xrd} and \ref{fig:lco_elec_prof}. 

Figure~\ref{fig:LCO_exp}B displays the \textit{ex-situ} Co \Ledge sXAS spectra collected from our \LCO thin film in total fluorescence yield (TFY) mode, with total electron yield (TEY) in Fig.~\ref{fig:lco_xas_supp}. Consistent with previous reports, charging leads to a small enhancement of the low-energy shoulder at 776~eV and a small energy shift of the main peak, 0.2~eV at 4.2~V and 0.4~eV at 4.8~V~\cite{degroot_lco,Mizokawa_LCO_XAS,Ensling_lco_band}. Such small changes are in sharp contrast to the strong metal \Ledge variations for cationic redox systems in non-layered compounds, \textit{e.g.} \LFPO~\cite{xiaosong_lfpo}. Figure~\ref{fig:LCO_exp}C shows the \textit{operando} sXAS spectra with the cycling voltage at representative values. Despite the high noise level that is typical for soft X-ray \textit{in-situ} tests, the spectral lineshapes evolve as in the \textit{ex-situ} data, with the main peak following the same trend (quantified in Fig.~\ref{fig:lco_edge_shift}). \textit{Ex-situ} O \Kedge sXAS measurements (Fig.~\ref{fig:LCO_exp}D) are consistent with previous reports~\cite{Mizokawa_LCO_XAS,Ensling_lco_band}. The strong pre-edge feature around 528~eV to 531~eV stems from the hybridized Co-O states with an extra low-energy feature growing upon charging~\cite{Mizokawa_LCO_XAS,Ensling_lco_band,subhayan_o_preedge}. While the \textit{operando} O \Kedge spectra display some interesting features at 533.3~eV (Fig.~\ref{fig:LCO_exp}E), careful re-evaluation shows that this new peak is from the organic solvent of the electrolyte, which suffers weak irradiation damage, leading to the peak intensity variation. Other \textit{operando} pre-edge features again follow the same trend as those \textit{ex-situ}. The consistency between \textit{ex-situ} and \textit{operando} data suggests that the small variation of Co \Ledge sXAS upon electrochemical cycling is not due to relaxation effects in \textit{ex-situ} experiments but is actually intrinsic to \LCO electrochemistry. This indicates that even at low voltages, delithiation does not follow a clearly defined \CoIII/\CoIV redox, which would lead to dramatic changes in the \Ledge lineshape.

\subsection{RIXS of \texorpdfstring{LiCoO$_2$}{LiCoO2}}

\begin{figure} 
	\centering
	\includegraphics[width=0.9\textwidth]{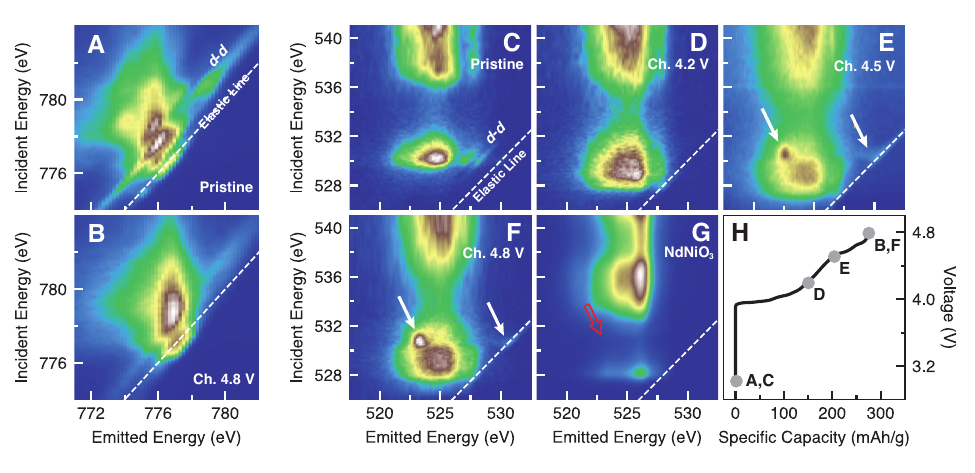} 
	\caption{\textbf{Experimental RIXS of \LxCO and \NdNO.}
	     (\textbf{A}) Co \Ledge RIXS of pristine \LCO. (\textbf{B}) Co \Ledge RIXS of \LxCO electrode charged to 4.8~V. (\textbf{C-F}) O \Kedge RIXS maps of \LxCO at representative charging voltages as indicated on the panels. Arrows in (\textbf{E, F}) indicate the oxidized oxygen features with O-O bond formation. (\textbf{G}) O \Kedge RIXS of a typical NCT system, \NdNO. The red arrow indicates the expected energy of RIXS features associated with O-O bond formation~\cite{zhou_o2_signature}. (\textbf{H}) Electrochemical profile of \LxCO and sampling points for \textit{ex-situ} RIXS. 
         Results of discharged electrodes after the 1st and 10th cycles are available in Fig.~\ref{fig:lco_k_rixs_extend}.}
	\label{fig:lco_rixs} 
\end{figure}

RIXS provides superior sensitivity to sXAS by characterizing excitations through the emission energy channel~\cite{tom_wanli_review}. Results at the Co \Ledge for pristine and charged electrodes are displayed in Fig.~\ref{fig:lco_rixs}A and~\ref{fig:lco_rixs}B, respectively. The most obvious change occurs between the well-defined \textit{d-d} excitations upon charging, indicated by the strong signals parallel to the elastic line in the pristine state, which largely disappear in the charged state. In addition, a stronger fluorescence in the charged state from the decay of valence band electrons implies an increase in covalency between Co and O upon charging.

Previously, RIXS signatures of oxidized anions were observed in highly charged \LCO around an excitation energy of 531~eV~\cite{Jienan_lco_doping}. However, this contrasted with neutron pair distribution function measurements that indicated no short O-O bond formation up to 4.6~V~\cite{Hu2021_LCO_RIXS}. Here, we performed \textit{ex-situ} RIXS of electrodes cycled to several representative voltages (Fig.~\ref{fig:lco_rixs}H). Up to 4.2~V (Fig.~\ref{fig:lco_rixs}D), RIXS shows only a broadening of features compared to the pristine state (Fig.~\ref{fig:lco_rixs}C). The characteristic features at an emission energy of 531 eV become clearly visible above 4.5~V, indicated by the two arrows in Fig.~\ref{fig:lco_rixs}E. The features remain strong at 4.8~V (Fig.~\ref{fig:lco_rixs}F) and display partial reversibility after 10 cycles (see Fig.~\ref{fig:lco_k_rixs_extend}).  

For a \LCO system without changes in composition or surface modifications, a notorious structural phase transformation takes place at around 4.25~V, with indications in electrochemical profiles (Fig.~\ref{fig:lco_elec_prof}). This transition triggers instabilities and irreversibility and has long been a central topic for optimizing the high-voltage performance of \LCO, with the focus on mitigating the structural transition~\cite{Jienan_lco_doping,liu_lco_doping}. It is interesting that the voltage where these oxidized oxygen features emerge in RIXS (Fig.~\ref{fig:lco_rixs}) coincides with the phase transformation. Nonetheless, no obvious change in O \Kedge RIXS could be seen below 4.2~V, other than the overall enhancement of the Co-O hybridization indicated by both O \Kedge and Co \Ledge RIXS. Therefore, questions remain about the redox mechanism at low voltages and whether the appearance of these RIXS features defines an onset of oxygen redox.

\subsection{Definition of anionic redox}

While debate remains on the nature of oxidized oxygen in layered cathodes~\cite{house_superstruct,House_o2,House_delocalO2,House_trappedo2,Tarascon_Peroxo,Seo2016,Kitchaev_pi_redox,Grey_LNO_Joule,ogley_metal_ligand,degroot_lco,Tsuchimoto_NMO,yang_cation_matter,Barbiellini_compton,Tarascon_NE_review,Gao_o2_cathode}, the two features in RIXS measurements, indicated by the arrows in Fig.~\ref{fig:lco_rixs}E-F, resemble signatures of strong O-O bonds in model systems~\cite{Zhou_RIXS_O2_CO2,zhou_o2_signature}. In particular, recent high-resolution RIXS measurements have revealed vibrational modes close to those of \Ogas molecules~\cite{house_superstruct}. Such findings have led to strong claims that the mechanism of oxygen redox corresponds to the formation and decomposition of trapped \Ogas molecules in many systems~\cite{house_superstruct,House_o2,House_delocalO2,House_trappedo2}, which may imply a sharp onset of oxygen redox above 4.25~V in \LCO. However, despite these strong claims, the model of trapped \Ogas has been challenged on both theoretical and technical grounds~\cite{Grey_LNO_Joule,Gao_o2_cathode}. A more recent study suggested that the features associated with \Ogas may emerge as extrinsic products of the X-ray measurement process in charged electrodes~\cite{Gao_o2_cathode, Wanli_NM2025}, making it all the more critical to re-evaluate the general relationship between oxidized oxygen and RIXS spectral features. 

It is crucial to emphasize that oxidized oxygen does not necessarily form chemical species with O-O bonding in solids. A typical example would be NCT systems that possess ground state wavefunctions with a large number of pre-existing ligand holes, \textit{i.e.} oxidized oxygen, as in \NdNO~\cite{Bisogni2016}. We investigated several prototypical NCT thin films -- $\mathrm{SmNiO_3}$, \NdNO, and $\mathrm{LaNiO_3}$ -- using O \Kedge RIXS, with the results shown in Fig.~\ref{fig:lco_rixs}G and Fig.~\ref{fig:ni_nct}A-B. These NCT systems do not possess the features associated with any chemically formed oxidized oxygen species. Additionally, highly covalent systems such as $\mathrm{CO_2}$, where electrons are largely shared between the two elements, do not have any features below an excitation energy around 533~eV in either RIXS (see Fig.~\ref{fig:ni_nct}C) or sXAS~\cite{Zhou_RIXS_O2_CO2}. Therefore, simply looking for those RIXS features would fail to detect the presence of oxidized oxygen states without O-O bonds. Thus, a key question emerges -- how to characterize oxides with a large number of oxygen holes.

\subsection{Numerical modeling of \texorpdfstring{LiCoO$_2$}{LiCoO2}}

Seminal work by Bisogni \textit{et al.} demonstrated the power of combined experimental and theoretical RIXS studies for revealing conclusive evidence of abundant oxygen holes in NCT systems~\cite{Bisogni2016}. Here, Figure~\ref{fig:lco_sim} shows RIXS and sXAS calculations for the Co \Ledge, using a charge transfer multiplet formalism~\cite{multiplet_code}, and the O \Kedge, using the {\sc ocean} Bethe-Salpeter Equation (BSE) code \cite{OCEAN,OCEAN3,QEspresso,ONCVPSP,r2scan,libxc}, of \LCO in the pristine/discharged and charged states. Technical details can be found in the Methods section and Supplementary Material. Theoretical RIXS maps and sXAS calculations (Fig.~\ref{fig:lco_sim}A-B) most consistent with both experiment (Fig.~\ref{fig:lco_rixs}A) and reference $\mathrm{LaCoO_3}$~\cite{Tomiyasu2017LaCoO3} measurements at the Co \Ledge yield a highly covalent, low-spin \dsix/\dsevenL ground state in the pristine material (Table~\ref{tab:lco_mult}), with a nominal \CoIII valence. O \Kedge calculations (Fig.~\ref{fig:lco_sim}C-D) also feature a low-spin Co configuration, consistent with the multiplet wavefunction and existing literature~\cite{van_Elp_LCO_1991,Van_der_Ven_old_LCO_DFT}. Combined, these observations place pristine \LCO at the boundary of NCT. 

Upon charging, multiplet calculations of the Co \Ledge reveal a transition to a purely NCT, low-spin ground state of mainly \dsevenLt character, with notable contributions from multiplet \dsixL and \deightLthree configurations (Table~\ref{tab:lco_mult}). Such a ground state wavefunction is only possible through increased covalency that stabilizes the ligand holes, and also results in broader excitations (Fig.~\ref{fig:lco_sim}E-F), consistent with experiment (Fig.~\ref{fig:lco_rixs}B). More notable changes occur at the O \Kedge after delithiation, as captured in our BSE calculations and experiment (Fig.~\ref{fig:lco_sim}G-H). The added hole density leads to rehybridization of the \eg manifold with the ligands orbitals (see Fig.~\ref{fig:lco_orb_pdos} and Table~\ref{tab:lco_mult}), which stabilize the added oxygen hole density, as noted above and first observed in prior LDA calculations~\cite{Van_der_Ven_old_LCO_DFT}. Taking this a step further, a Bader charge analysis~\cite{bader} comparing the lithiated and delithiated structures shows a loss in charge primarily on oxygen (Table~\ref{tab:lco_bader}).  Our simulations, combining observations from both the Co $L_3-$ and O \Kedge, confirm the picture of a transition to a pure NCT ground state, which naturally provides the channel for anionic redox through enhanced oxygen hole states.

\subsection{Experiments and theory of \texorpdfstring{LiNiO$_2$}{LiNiO2}}

We performed complementary experimental and theoretical studies on another layered oxide, \LNO, to test the prevalence of NCT evolution in cathode electrochemistry. Fig.~\ref{fig:lno_sim}A,D and Fig.~\ref{fig:nio_rixs} show the experimental Ni \Ledge RIXS maps of pristine \LNO, charged \LxNO, and NiO, respectively. As the nominal valence of Ni increases, features at low incident energy weaken around 852.7~eV  while high energy features are strongly enhanced around 853~eV to 856~eV. These features and the evolution of their intensity correspond to the typical two peaks in sXAS, consistent with previous literature~\cite{Wanli2019_LNO_RIXS}. However, RIXS reveals far richer information. The low-energy peak in sXAS is dominated by a strong {\it d-d} excitation in RIXS, whereas the high-energy peak in sXAS corresponds to fluorescence signals of a completely different nature and profile in RIXS. This is indicative of evolution toward a highly covalent state upon charging, just like \LCO, and the RIXS and sXAS profiles of charged \LNO resemble the spectra for \NdNO~\cite{Bisogni2016}, with an unambiguous NCT ground state enhanced upon charging.

Theoretically, \LNO is a “high-entropy charge glass”,\cite{Kateryna2019disp} complicating calculations that require a superposition of electronic states from a Jahn-Teller distorted and size disproportionated \Nioct octahedra (See Methods). Taking this into account, our calculations at the Ni \Ledge(Fig.~\ref{fig:lno_sim}B,C) point to an appreciable number of oxygen holes in the ground state wavefunction of pristine \LNO, formed from a combination of states for the three octahedral environments with majority \deight, \deightL, and \deightLt character, respectively (Table~\ref{tab:lno_mult}). Simulated O \Kedge BSE spectra (Figs.~\ref{fig:lno_ok_rixs}A-C) similarly show distinct contributions from different oxygen bonding environments (Figs.~\ref{fig:lno_orb_pdos}) in sXAS and reproduce the general shape of the experimental RIXS. Combined, these observations already place pristine \LNO squarely in a NCT state. 

Unlike \LCO, the presence of a NCT ground state in pristine \LNO leads to significant changes at the Ni \Ledge and less notable evolution of the O \Kedge spectra upon charging, even though the average covalency increases substantially. The increase in the intensity of the Ni \Ledge sXAS and RIXS high energy peak in charged \LxNO (Fig.~\ref{fig:lno_sim}E,F) originates from an increased population of the compressed \Nioct, with a majority \deightLt configuration (Table~\ref{tab:lno_mult}). This is also reflected in the narrowing of the O \Kedge pre-edge peak near 528.5~eV (Fig.~\ref{fig:lno_ok_rixs}D-F). Again, a Bader charge analysis~\cite{bader} shows a loss in charge primarily on oxygen upon delithiation (Table~\ref{tab:lno_bader}). Although more complex, our combined experimental and theoretical analysis at both the Ni $L_3-$ and O \Kedge reveals that charging of \LNO is also facilitated through the enhancement of NCT states with increasing ligand holes, mirroring the overall trends in the \LCO.

\section{Conclusion}\label{sec:conclusion}

Our analysis of \LCO, \LNO, and other model systems suggests a universal redox mechanism in high-energy layered compounds. Instead of the widely accepted low-voltage-cationic/high-voltage-anionic model, we show that enhancement of the NCT ground state upon charging provides the fundamental pathway for redox throughout the whole voltage range. The NCT ground state is strongly hybridized, however, is distinct from conventional TM-O hybridization that does not offer ligand hole states in systems like \LFPO. Prior quantitative sXAS studies on \LFPO~\cite{xiaosong_lfpo}, and RIXS studies in Fig.~\ref{fig:lfpo_rixs}, show well-defined and localized \textit{d-d} excitations at the Fe \Ledge parallel to the elastic line in both the discharged and charged states, with the substantial differences between them reflecting true cationic redox: \FeII to \FeIII. Although charging also enhances Fe-O hybridization~\cite{subhayan_o_preedge}, these localized RIXS features are in contrast to the broad and featureless RIXS profile of charged \LCO and \LNO, indicating an unusually strong hybridization in NCT states.

We could now categorize the redox activities in battery cathodes into three types. First, a TM-based redox center with conventional TM-O hybridization, such as \LFPO. A practical limitation of this redox type is the relatively low voltage regulated by the conventional range of positive charge transfer states~\cite{green_sawatzky_NCT}. Second, in layered oxides that could potentially operate at high voltages, the NCT state emerges from the very beginning of charging, and the NCT enhancement itself is the redox channel through its inherent ligand hole states. The unusually covalent NCT state means that there is no literal atomic redox center. The highly covalent and itinerant nature of this state could be the fundamental reason for the facile kinetics and high reversibility of \LCO electrodes in the voltage range used in commercial batteries. Importantly, O-O bonds do not form in this scenario although an equivalent number of oxygen holes exist in the lattice.

Third, at higher voltages, \textit{e.g.} above 4.25~V for \LCO, the increase of ligand hole density reaches a critical level where highly oxidized oxygen tends to get close or even form O-O bonds to reduce the total energy, which could be the fundamental driving force behind the structural phase transitions at high voltage. Continuing to increase the voltage eventually leads to over-enhanced NCT, \textit{i.e.} excess ligand holes, leading to literal O-O bond and detrimental \Ogas and radical oxygen formation. To resolve the debate on oxygen redox at high voltages, it is crucial to understand that the formation of \Ogas is fundamentally due to over-enhanced NCT and excess ligand holes, and thus could be triggered by influences other than electrochemistry. For example, it was recently suggested that X-ray photons could weaken the TM-O bonding states leading to the formation of  \Ogas~\cite{Gao_o2_cathode}. With external triggers like X-rays, the system could form \Ogas before its formation voltage in electrochemistry. This explains the common observation of \Ogas features in RIXS at a lower voltage than for electrochemical gas evolution~\cite{Oswald_stability_2023}. It is also well known that charging at high-temperature could also trigger more \Ogas or radical oxygen formation leading to electrode/electrolyte instability issues. Therefore, electrochemical voltage, temperature, and X-rays all represent extrinsic triggers for O-O bond formation, which could only evolve from a system with intrinsic oxygen holes in NCT states\cite{Wanli_NM2025}. 

The practical ranges of these three types of redox behaviors are obviously material dependent; however, they are based on the same fundamental parameter -- ligand hole density in the NCT states. This is consistent with a prior proposal suggesting that the number of oxygen holes could be a critical index for qualitative transitions between reversible and irreversible redox behavior~\cite{Doublet_NM}. This also sets the fundamental difference between \LCO and \LNO, corresponding to their different electrochemical behaviors, as pristine \LNO, already in a NCT state, contains a significant amount of ligand holes.

This NCT-based redox mechanism is key to understanding electrochemical performance vis-\`{a}-vis voltage, stability, and kinetics. It unifies the understanding of seemingly contradictory observations of structural transitions, voltage hysteresis, oxygen dimerization, and gas release without a conventional redox center. Harnessing and characterizing the NCT state without triggering O-O bond formation could be the key to access viable high-voltage cycling and could be achieved by enhancing overall covalency through elemental doping, \textit{e.g.} F or $4d$/$5d$ TMs, something emphasized in very early work~\cite{Sathiya2013,Grimaud2016}, but quickly overlooked. This study also reveals a vast family of NCT systems for further fundamental physics studies, which will lead to more interdisciplinary discussions and collaborations between fundamental and practical fields towards innovative materials for high-energy batteries.

\pagebreak
\section{Figures}\label{sec:figs}

\begin{figure} 
	\centering
	\includegraphics[width=0.9\textwidth]{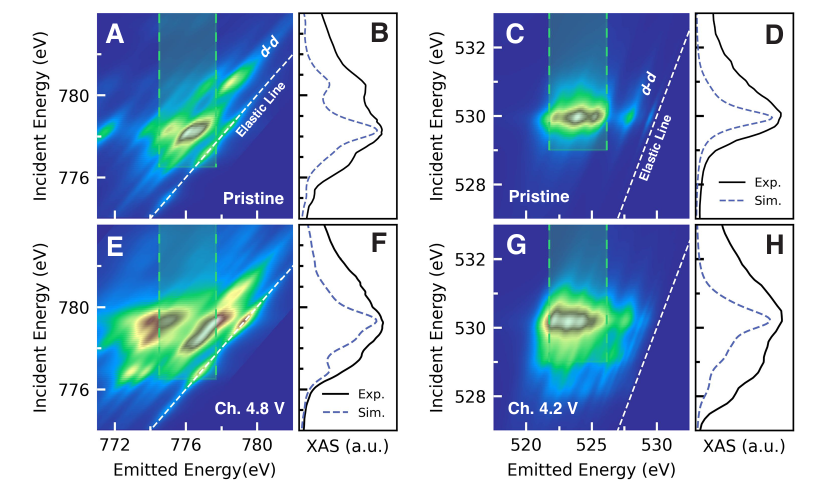} 
	\caption{\textbf{sXAS and RIXS calculations of LiCoO\textsubscript{2}.}
		(\textbf{A-B}) Co \Ledge RIXS and sXAS calculations of pristine \LCO. (\textbf{C-D}) O \Kedge RIXS and sXAS of pristine \LCO. (\textbf{E-F}) Co \Ledge RIXS and sXAS of charged \LxCO. (\textbf{G-H}) O \Kedge RIXS and sXAS of charged \LxCO. sXAS comparisons with experiment in panels \textbf{B}, \textbf{D}, \textbf{F}, and \textbf{H}. RIXS experiments in Fig.~\ref{fig:lco_rixs}, emission lines overlaid in theoretical plots as guides to the eye.}
	\label{fig:lco_sim} 
\end{figure}

\begin{figure} 
	\centering
	\includegraphics[width=0.9\textwidth]{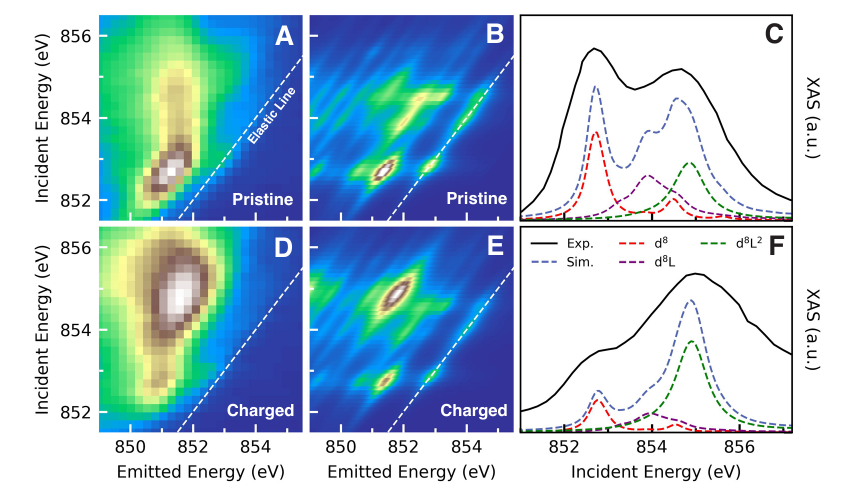} 
	\caption{\textbf{Experiments and theory of LiNiO\textsubscript{2}.}
      Experimental (\textbf{A}) and simulated (\textbf{B}) Ni \Ledge RIXS and sXAS (\textbf{C}) of pristine \LNO. Experimental (\textbf{D}) and simulated (\textbf{E}) Ni \Ledge RIXS and sXAS (\textbf{F}) of charged \LxNO.}
	\label{fig:lno_sim} 
\end{figure}

\backmatter

\pagebreak
\section*{Supplementary information}

Details of experiments and theoretical calculations are available in the supplementary file.

\section*{Acknowledgements}

\LNO material and samples are prepared at Argonne National Laboratory supported by the Office of Vehicle Technologies of the U.S. Department of Energy. \LCO material and \textit{ex-situ} electrodes are produced at the US DOE CAMP (Cell Analysis, Modeling and Prototyping) Facility, supported by the DOE VTO at Argonne National Laboratory. Thin films of Nd/Sm/La-NiO\textsubscript{3} are synthesized at Tsinghua University with support from The National Natural Science Foundation of China (No. 52025024).

Theoretical work is supported by the U.S. Department of Energy, Office of Basic Energy Sciences, Materials Sciences and Engineering Division. Calculations were performed on the Sherlock cluster at Stanford University and on resources of the National Energy Research Scientific Computing Center (NERSC), a Department of Energy Office of Science User Facility, using NERSC award BES-ERCAP0031424.

Spectroscopic experiments are performed at Beamline 8.0.1.1 of Advanced Light Source, which is a DOE Office of Science User Facility under Contract No. DE-AC02-05CH11231.

\section*{Declarations}

Certain equipment, instruments, software, or materials are identified in this paper in order to specify the experimental procedure adequately.  Such identification is not intended to imply recommendation or endorsement of any product or service by NIST, nor is it intended to imply that the materials or equipment identified are necessarily the best available for the purpose.

\section*{Author contribution}
T.P.D. and W.Y. conceived the study. E.G.L and K.H.H. conducted multiplet calculations and analysis. K.H.H. and L.I. conducted DFT and {\sc ocean} calculations and performed the subsequent analysis.. J.J.K. conducted EXAFS calculations. W.Y. supervised interpretation of XAS and RIXS measurements. J.V., J.J.K. and J.R. advised {\sc ocean} and calculation set-up and interpretation. Q.L. and S.Y. synthesized \LCO thin films. B.J.P. synthesized and fabricated \LCO electrodes. Y.W., H.L. and P.Y. synthesized NCT thin films. J.G., B.S. and E.L. synthesized and fabricated \LNO electrodes. Q.L., G.H.L., Z.Z., J.W., Z.X.S., Y.S.L., J.H.G., Y.D.C. and W.Y. performed the XAS and RIXS experiments. E.G.L, K.H.H., B.M., T.P.D., and W.Y. wrote the manuscript, and all authors revised the manuscript. 

\section*{Competing interests} There are no competing interests to declare.

\nocite{*}
\bibliography{sn-bibliography}

\clearpage
\newpage

\setcounter{figure}{0}
\clearpage
\newpage

\renewcommand{\thefigure}{S\arabic{figure}}
\renewcommand{\thetable}{S\arabic{table}}
\renewcommand{\theequation}{S\arabic{equation}}
\renewcommand{\thepage}{S\arabic{page}}
\setcounter{figure}{0}
\setcounter{table}{0}
\setcounter{equation}{0}
\setcounter{page}{1}

\section*{Supplement}

\subsection*{Materials and Methods}

Certain equipment, instruments, software, or materials are identified in this paper in order to specify the experimental procedure adequately.  Such identification is not intended to imply recommendation or endorsement of any product or service by NIST, nor is it intended to imply that the materials or equipment identified are necessarily the best available for the purpose.

\subsubsection*{LiCoO\textsubscript{2} thin film growth}

Thin film \LCO was prepared on stainless steel substrate and $\mathrm{Si_3N_4}$ window by pulsed laser deposition (PLD). The base pressure in the deposition chamber was below $\mathrm{2\times10^{-5}}$~Pa. The \LCO target was sintered with 15\% excess $\mathrm{Li_2O}$ to compensate for the Li loss during deposition. The target was ablated by KrF excimer laser (wavelength of 248~nm) with the laser pulse repetition rate of 2~Hz and laser energy of 300~mJ/pulse. The samples were grown at 500 centigrade with 10 Pa oxygen pressure to obtain high temperature phase \LCO film with a thickness of $\approx$150~nm. Then, the deposited films were cooled down to room temperature within the chamber under the same oxygen pressure.

\subsubsection*{\textit{in-situ/operando} LiCoO\textsubscript{2} cell assembly}
The \textit{in-situ} static cells were assembled in an argon glove box. The static cell utilizes a thin $\mathrm{Si_3N_4}$ (100~nm) membrane to separate ultrahigh vacuum atmosphere of the chamber and well-sealed liquid electrolyte of the cell. The ultrathin $\mathrm{Si_3N_4}$ window allows incident synchrotron radiation into the cell and emitted X-rays out. The \LCO thin film deposited on $\mathrm{Si_3N_4}$/Au membrane was utilized as the cathode and Li foil as the counter and reference electrode, commercial $\mathrm{LiPF_6}$ in EC:EMC:DEC (4:2:4) was well sealed in the static cell as the liquid electrolyte. The assembled static cell was then transferred into the loadlock of spectroscopic system for gentle pump down, then into the main experimental chamber.

\subsubsection*{LiCoO\textsubscript{2} Swagelok cell assembly and electrochemical characterizations}
In order to prepare cycled \LCO electrodes for RIXS experiments, we prepared Swagelok cells with standard \LCO electrodes from the U.S. Department of Energy’s (DOE) CAMP (Cell Analysis, Modeling and Prototyping) Facility, Argonne National Laboratory. Li-ion batteries with \LCO cathode were assembled in Swagelok cells for electrochemical characterization. Li foils were used as the counter and reference electrode, and the same electrolyte was used as before (LiPF$_6$ in EC:EMD:DEC). 
Battery assembly was carried out in argon filled glove box with H\textsubscript{2}O and \Ogas content less than 1~ppm. The galvanostatic charge/discharge experiments were conducted at room temperature in a voltage range between 3.0~V to 4.8~V at a rate equal to the capacity divided by 10 hours (``C/10'' rate). 

\subsubsection*{Perovskite nickelates thin film growth}
Thin films of $\mathrm{NdNiO_3, SmNiO_3, LaNiO_3}$ were synthesized using a custom-designed pulsed laser deposition (PLD) system. The depositions were carried out at $600^]\circ$~C under an oxygen partial pressure of 10~Pa. A KrF excimer laser ($\lambda$ = 248~nm) with an energy density of 2.0~J\,cm$^{-2}$ and a repetition rate of 2~Hz was used to ablate the stoichiometric ceramic targets. Following growth, the samples were cooled to room temperature at a cooling rate of $10^\circ$~C per minute under the same oxygen pressure. Structural characterization by X-ray diffraction (XRD) confirmed the high crystallinity of the films.

\subsubsection*{Preparation of Delithiated LiNiO\textsubscript{2} Samples for ex situ Ni \textit{L}\textsubscript{3}-edge RIXS}

The \LNO electrodes, prepared at Argonne National Laboratory, consisted of \LNO powder synthesized under the optimized condition from the effort of Realizing Next Generation Cathode (RNGC) consortium project, a carbon additive (C-45P), and a poly(vinylidene fluoride) (PVdF) binder on an aluminum foil substrate. For delithiation, CR2032-type coin cells were assembled using the \LNO electrode as the working electrode, a lithium metal disc as the counter electrode, a Celgard 2320 separator, and a commercial electrolyte of 1.2~M LiPF6 in ethylene carbonate (EC) and ethyl methyl carbonate (EMC) (3:7, wt. mixture). The cells were charged using a constant current/constant voltage (CC/CV) protocol; a constant current of ``C/0.01'' rate (1C = 200~mA\,g$^{-1}$) was applied until the voltage reached 4.8~V, followed by a constant voltage hold at 4.8~V until the current tapered to ``C/0.01'' rate. Immediately following the delithiation process, the cells were disassembled inside an argon-filled glovebox, and the retrieved electrode was washed with dimethyl carbonate (DMC).

\subsubsection*{Soft X-ray spectroscopy}
\textit{In-situ} and \textit{ex-situ} Soft X-ray XAS and RIXS experiments were performed at the iRIXS and wetRIXS endstation at Beamline 8.0.1 of the Advanced Light Source at Lawrence Berkeley National Laboratory. All electrode samples were prepared in an Argon glove box with water and oxygen concentrations below 1~ppm. The cycled electrodes were extracted from the Swagelok cell after cycling the desired voltages, and washed with dimethyl carbonate (DMC) in the Ar glove box. The electrodes were then cleaved from their current collector and mounted on a sample holder. All samples were transferred into measurement chamber by using a home-made sample transfer kit to avoid any air exposure. 

The sXAS experimental resolution is 0.15~eV (full-width gaussian) without considering core-hole broadening, which is about 0.2~eV. The sXAS data shown in this work are measured in both total fluorescence yield (TFY) and total electron yield (TEY) mode with probe depth of about 100~nm and 10~nm, respectively. The absolute values of excitation energies of O \Kedge spectra in this work were calibrated by measuring a reference anatase TiO\textsubscript{2} and set the lowest energy peak at 530.75~eV. The energy calibrations for transition metal $L-$edges were based on our previous benchmark works~\cite{qinghao_xray_redox}.

RIXS measurements were performed at iRIXS endstation at Beamline 8.0.1 of the ALS~\cite{iRIXS}. Data were collected through the high-efficiency modular spectrometer~\cite{sXAS_rev}. The resolution of the excitation energy is about 0.2~eV, and the emission energy about 0.25~eV. An excitation energy step size of 0.2~eV was chosen for all the RIXS maps. It takes about one minute to collect the RIXS of each excitation energy and the final 2D RIXS images are obtained after normalizations to the beam flux and collection time. In order to reduce the irradiation effect on the signals, the system allows a largest possible beam size of about 25~µm by 150~µm to achieve the resolution mentioned above. During the data collection, the sample was manipulated in a rastering mode to maintain a constant movement under X-rays.

\subsubsection*{Multiplet calculation} \label{sec:multiplet}

To model the core level spectroscopy of \LxCO and \LxNO, we exactly diagonalize a full atomic multiplet including charge transfer and hybridization effects of a transition metal center (5 $d$ orbitals), 3 oxygen ligands (3 $p$ orbitals each). The single site is centered at (\begin{math}\pi/2,\pi/2\end{math}) in momentum~\cite{multiplet_code}. The Hamiltonian for the multiplet cluster can be expressed as:
\begin{align}
    \hat H &= 
     \frac{1}{2}\sum_{i,\sigma,\sigma'} \sum_{\mu,\nu,\mu',\nu'} U_{\mu,\nu,\mu',\nu'} \hat c^\dagger_{i,\mu,\sigma}\hat c^\dagger_{i,\nu,\sigma'}\hat c_{i,\mu',\sigma'}\hat c_{i,\nu',\sigma} \nonumber\\ 
     & + \sum_{i,j,\sigma}\sum_{\mu,\nu} t_{i,j}^{\mu,\nu} \hat c^\dagger_{i,\mu,\sigma}\hat c_{j,\nu,\sigma} \nonumber + \sum_{i,\mu,\nu,\sigma}V_{CEF}(\mu,\nu)c^\dagger_{i,\mu,\sigma}\hat c_{i,\nu,\sigma}\nonumber \\
& + \frac{1}{2}\sum_{i,\sigma,\sigma'} \sum_{\mu,\nu,\mu',\nu'} U_{\mu,\nu,\mu',\nu'} \hat c^\dagger_{i,\mu,\sigma}\hat d^\dagger_{i,\nu,\sigma'}\hat c_{i,\mu',\sigma'}\hat d_{i,\nu',\sigma} \nonumber\\
& - \sum_{i,\sigma,\sigma'} \sum_{\mu,\nu} \lambda_{\mu,\nu}^{\sigma,\sigma'} \hat d^\dagger_{i,\mu,\sigma} \hat d_{i,\nu,\sigma'} + \sum_{i}\Delta_{i}n_{i}
\end{align}

Where $i$, $j$ refer to the different atomic sites, $\mu$, $\nu$ refer to different sets of $l$, $m$ quantum numbers, and $\sigma$ refers to spin. The creation and annihilation operators denoted in $c$ are valence electron (hole) operators; whereas operators denoted in $d$ are core electron (hole) operators. 

The first term includes a Hubbard-like U term for the coulomb direct and exchange interactions for TM oxides. The second term includes a \textit{t} hopping element between different atomic sites and their orbitals. The third term includes an octahedral crystal field ($V_{CEF}$) for the d-orbtials in the metal atom, the fourth term is the core-valence coulomb interaction, the fifth term is the spin-orbit coupling $\lambda$ at the core and the last term is the charge transfer energy $\Delta$ at each atomic site. The multi-particle eigenstates for a Hamiltonian of an \textit{N} hole cluster and one for a \textit{N-1} hole cluster with a core hole serve as the initial (\textit{i}), intermediate ($\nu$), and final (\textit{f}) states for the calculation of XAS by Fermi's golden rule:
\begin{align}
&\kappa_{e_i, k_i}(\omega)=
\frac{1}{\pi Z} \sum_{i,\nu} e^{-\beta E_i} \mid \langle\nu\mid \hat D_{k_i}({e_i})\mid i \rangle\mid^2 \delta(\omega-(E_\nu-E_i))
\end{align}
And for RIXS using the Kramers-Heisenberg representation:
\begin{align}
R( e_i,e_f,k_i,k_f,\omega_i,\omega_f)=&\frac{1}{\pi Z} \sum_{i,f} e^{-\beta E_i} \left\vert \sum_\nu \frac{\langle f \mid \hat D_{k_f}^*(e_f)\mid\nu\rangle \langle\nu\mid \hat D_{k_i}({e_i})\mid i\rangle}{\omega_i-(E_\nu-E_i)-i \Gamma}\right\vert^2 \nonumber \\
&\delta(\Omega-(E_f-E_i))
\end{align}
Where \begin{math}E_{i,\nu,f}\end{math} refers to the eigenenergy and \begin{math} D_{k_i}({ e_i})\end{math} is the dipole operator for a photon of frequency $\omega$, momentum \textit{k} and polarization \textit{e}, and $\Omega=\omega_i-\omega_f$. To account for finite lifetime effects, a Lorentzian broadening was applied to the spectral function. For transition metal \Ledge, a broadening with half width at half maximum (HWHM) of 0.5 eV is applied in the absorption energy, while a broadening of 0.3 eV is applied in the loss energy axis.


\subsubsection*{Technical details for the multiplet calculations}

The parameters, resulting orbital occupations and configuration interaction ground states can be found in Table~\ref{tab:lco_mult} and~\ref{tab:lno_mult} for Co and Ni calculations, respectively.

Concerning \LCO: Previous work~\cite{van_Elp_LCO_1991} has noted that the lower incident energy features (775~eV to 777~eV) arise from a cobalt oxide contaminant with a lower formal oxidation state of \CoII. Hence, we separately model both systems and overlay a \CoII and \CoIII cluster with a 1:4 ratio to accurately reproduce the experimental data (Fig.~\ref{fig:lco_sim}A-B). For the charged case, since \LxCO is partially charged ($\mathrm{x > 0}$), the charged material still exhibits some signal from the lithiated case. Hence, for our total charged calculations, we add a \CoIV calculation to the pristine calculation in a 2:1 ratio (Fig.~\ref{fig:lco_sim}E-F). All Co \Ledge spectra have been shifted by 781.4 eV to fit experimental measurements.

Concerning \LNO: For the pristine \LNO calculation, due to the “charge-glass” like structure~\cite{Kateryna2019disp}, we combine calculations of \NiIII, \NiIII and \NiIV clusters with a 1:1:1 ratio, which reproduces the experimental data (Fig.~\ref{fig:lno_sim}A-C). Similar to \LxCO, the charged material of \LxNO also exhibits some signal from the lithiated case. Here, we adjust the ratio of the calculations for \NiIII, \NiIII and \NiIV clusters to a 1:1:4 ratio, accounting for 25\% of the remaining pristine material (Fig.~\ref{fig:lno_sim}D-F). All Ni \Ledge spectra have been shifted by 857.45~eV to fit experimental measurements.

\subsubsection*{DFT ground state calculation} \label{sec:dft}

\begin{figure} 
	\centering
	\includegraphics[width=0.9\textwidth]{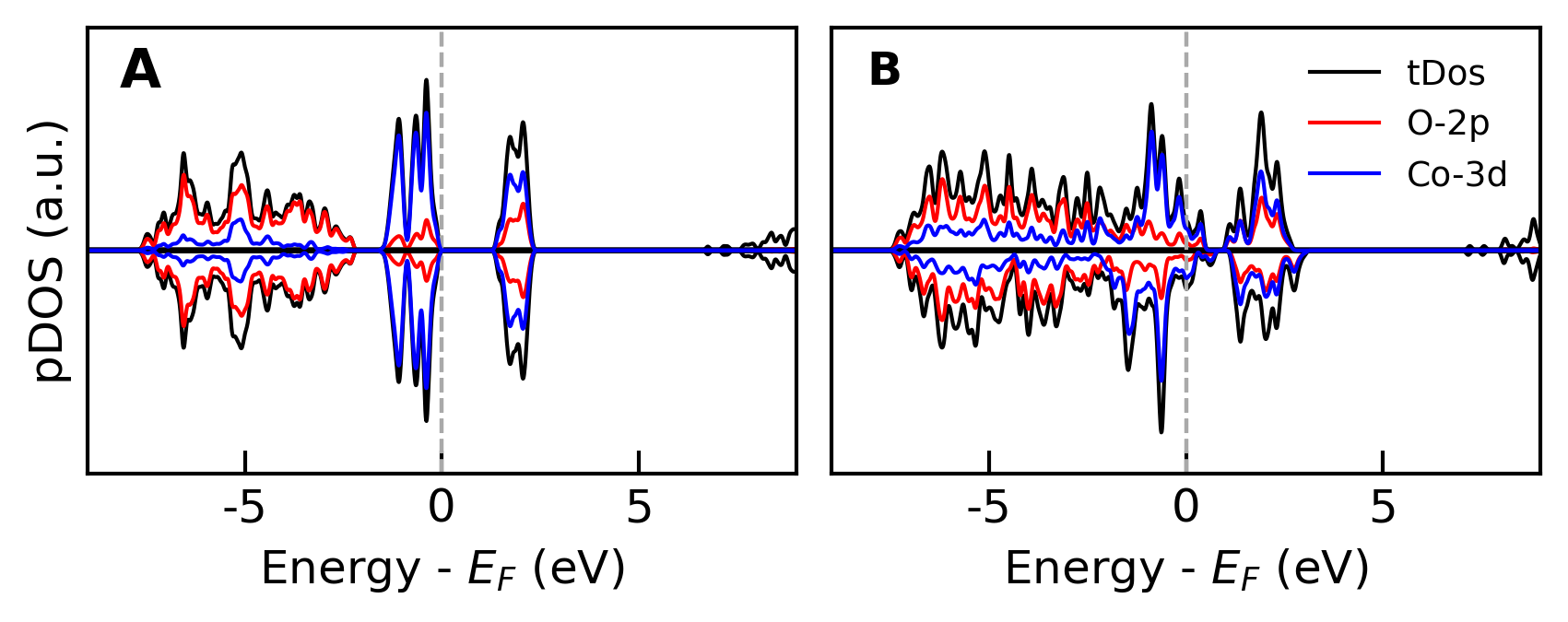} 

	\caption{\textbf{Density of states for \LxCO.} (\textbf{A}) Total density of states (tDOS) and partial density of states (pDOS) for \LCO. The Fermi energy is represented by the dashed grey line. (\textbf{B}) pDOS for \LqCO. The weight of the oxygen band above the Fermi level increases upon lithium removal, indicating that the charge compensation is highly covalent and involves both cobalt and oxygen atoms. The pDOS qualitatively correlate to the change observed in the oxygen \Kedge sXAS.}
	\label{fig:lco_pdos} 
\end{figure}

\begin{figure} 
	\centering
	\includegraphics[width=0.8\textwidth]{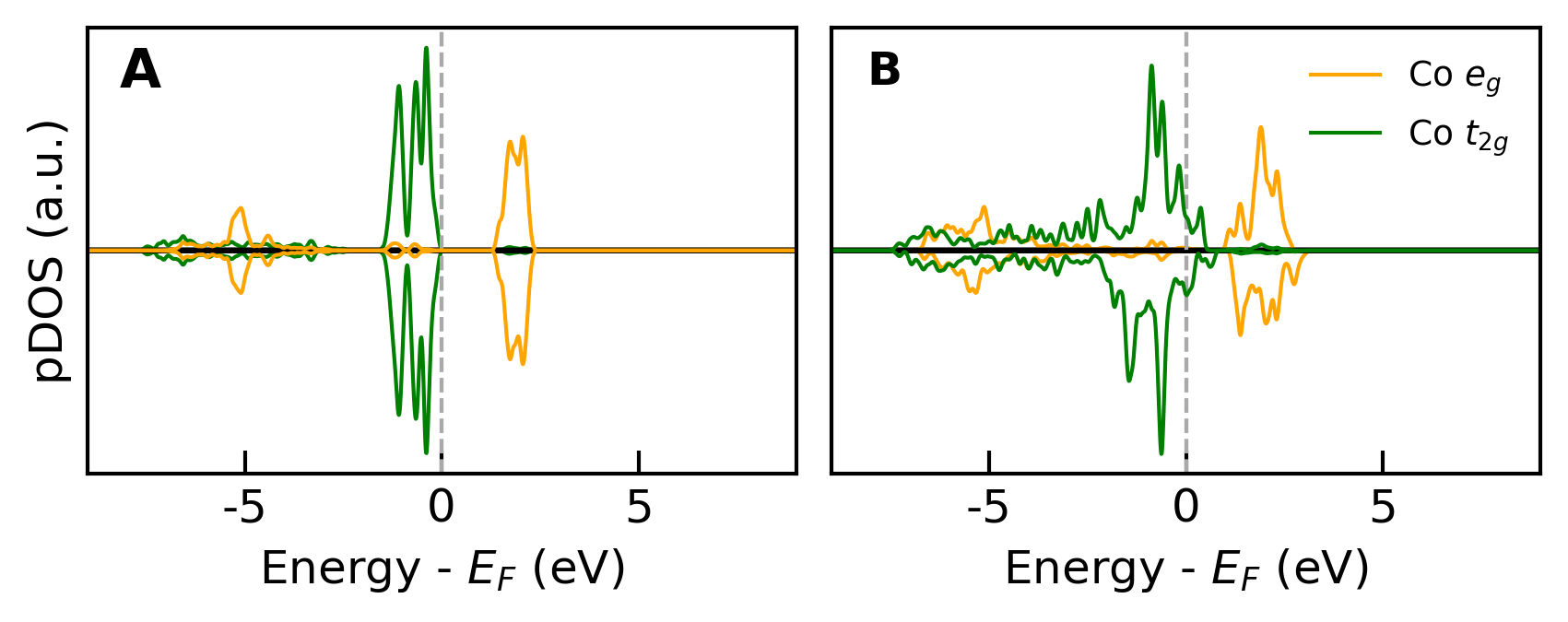} 

	\caption{\textbf{Projected Density of states for \LxCO onto Co-3$d$ manifold.} (\textbf{A}) Partial density of states projected onto Co-3$d$ manifold for \LCO and (\textbf{B}) \LqCO. Rehybridization between the Co-3$d$ and O-2$p$ orbitals can been seen through the change in the Co d\textsubscript{xy} orbitals, consistent with our multiplet calculation and previous reports~\cite{Van_der_Ven_old_LCO_DFT}.}
	\label{fig:lco_orb_pdos} 
\end{figure}

\begin{figure} 
	\centering
	\includegraphics[width=0.8\textwidth]{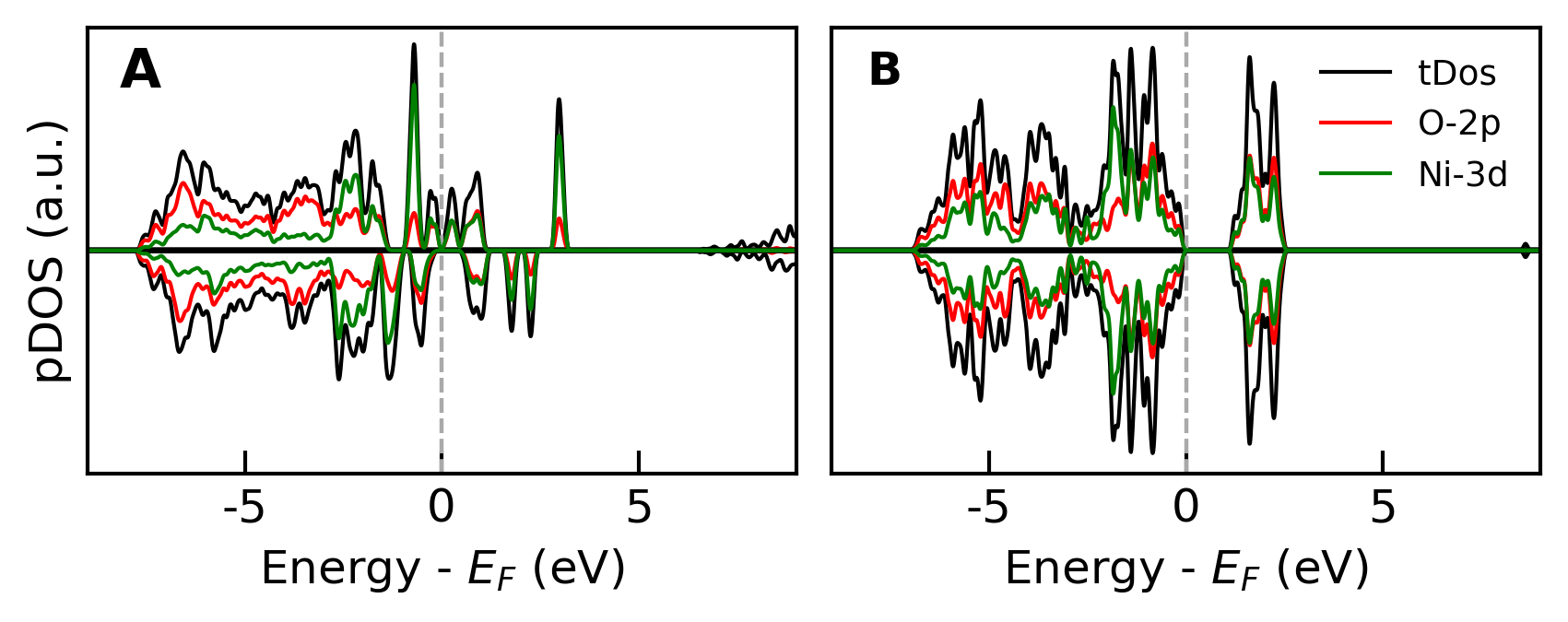} 

	\caption{\textbf{Density of states for \LxNO.} (\textbf{A}) Total density of states (tDOS) and partial density of states (pDOS) calculated with the 67\% SD structure. The Fermi energy is represented by the dashed grey line. The calculations results in a 0.2~eV band gap. The conduction band near the fermi level ($\mathrm{E-E_F}<$~1~eV) is associated with the small \Nioct (\deightLt) octahedra. The conduction band around 2~eV above the fermi level is associated with the Jahn-Teller distorted \Nioct (\deightL) octahedra, whereas the oxygen band 3.5~eV above the fermi level is associated with the big \Nioct (\deight) (b) tDOS and pDOS for $O3$ stacked \NO, here the weight of oxygen bands also increases significantly in the conduction band upon lithium removal.} 
	\label{fig:lno_pdos} 
\end{figure}

\begin{figure} 
	\centering
	\includegraphics[width=0.5\textwidth]{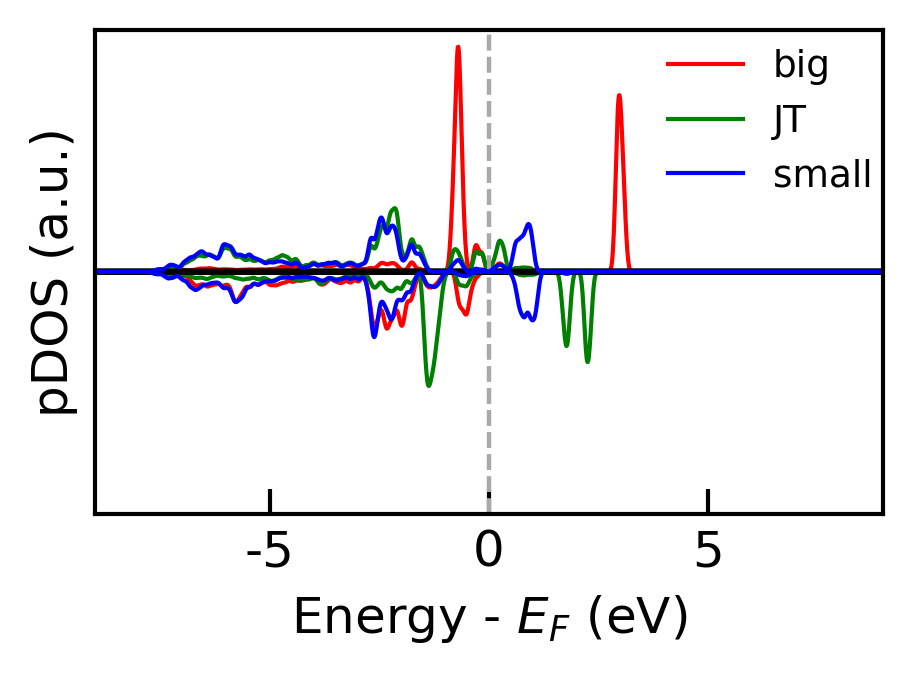} 

	\caption{\textbf{Projected density of states on to Ni-3\textit{d} manifold calculated with the 67\% SD LiNiO\textsubscript{2} structure.} Different \Nioct bonding environment have distinct contributions to the conduction density of state, which results in the complex pre-edge structure seen in the O \Kedge sXAS and RIXS.}
	\label{fig:lno_orb_pdos} 
\end{figure}

\begin{figure} 
	\centering
	\includegraphics[width=0.9\textwidth]{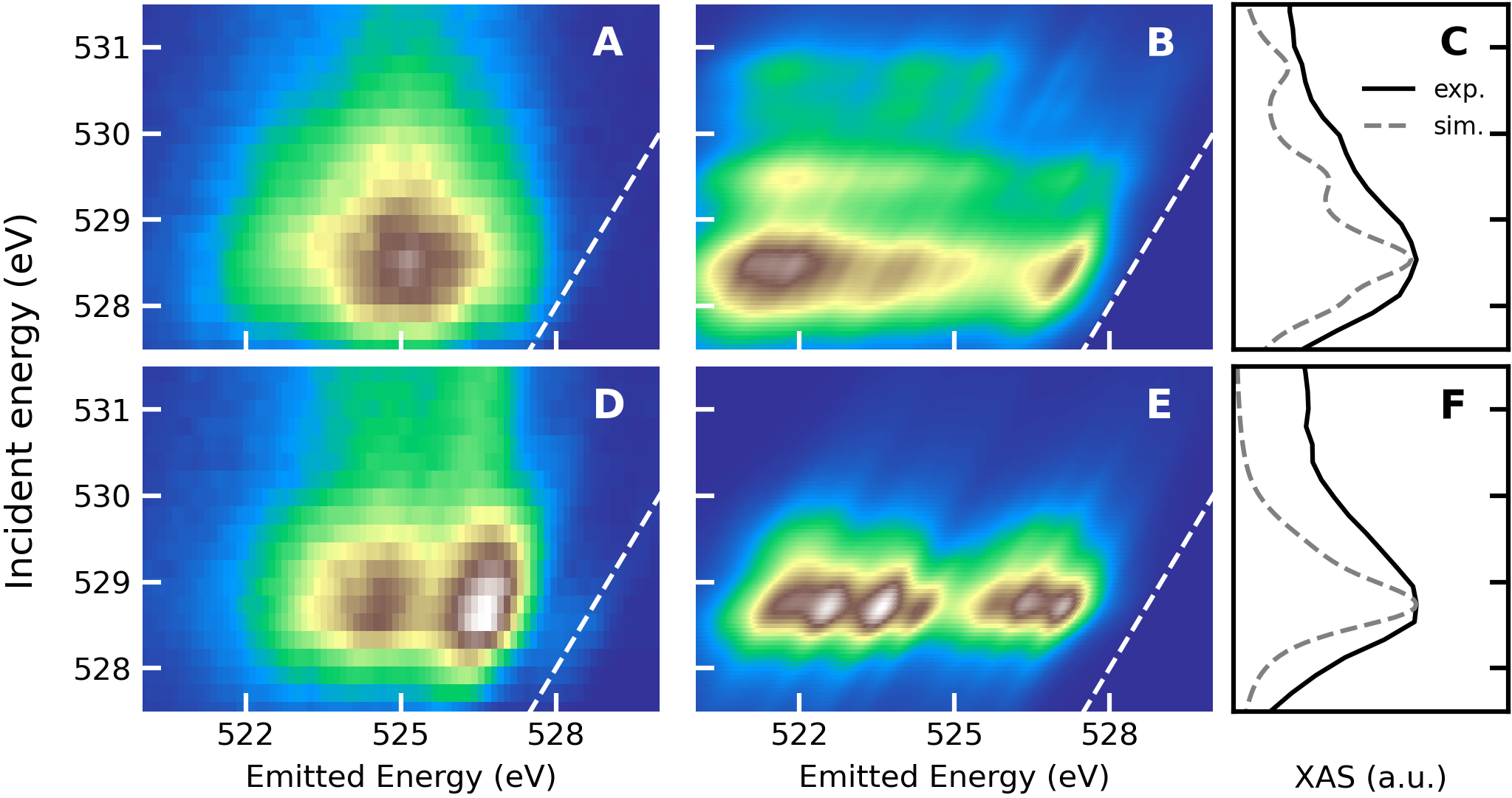} 

	\caption{\textbf{O \textit{K}-edge sXAS and RIXS for \LxNO.} (\textbf{A}) Experimental O \Kedge RIXS of \LNO. (\textbf{B}) {\sc ocean} simulated O \Kedge RIXS using 67\% SD \LNO structure. (\textbf{C}) Experimental and simulated O \Kedge sXAS of \LNO. (\textbf{D}) Experimental O \Kedge RIXS of charged \LxNO. (\textbf{E}) {\sc ocean} simulated O \Kedge RIXS using $O3$ stacked \NO. (\textbf{F}) Experimental and simulated O \Kedge sXAS of \LxNO. The simulated RIXS spectra are broad in the energy loss axis, likely due to the de-localized valence bands shown in Fig.~\ref{fig:lno_pdos}.
    }
	\label{fig:lno_ok_rixs} 
\end{figure}

All DFT ground state calculations were done using the QuantumEspresso DFT package~\cite{QEspresso}. We generated pseudopotentials using the ONCVPSP code~\cite{ONCVPSP} (version 3.3.1) with PBE parametrization. Fixed cell relaxation is performed with the r2SCAN functional~\cite{r2scan} implemented by the libxc library~\cite{libxc}, as meta-GGA functionals are reported to have good performance on layered oxide materials~\cite{Chakraborty2018}. The force convergence criterion is set to $ <10^{-4}$~Ry for all force components with a plane-wave cutoff of at least 140~Ry. All structures are calculated with the magnetic ordering on the transition metals that has the lowest ground-state energy, using a gamma-centered k grid with a linear density of at least 0.2~$\mathrm{Bohr^{-1}}$ using Gaussian smearing with $\mathrm{\sigma=0.02}$~eV. Bader charge~\cite{bader} (Table~\ref{tab:lco_bader}-\ref{tab:lno_bader}) and projected density of states (pDOS) (Fig.~\ref{fig:lco_pdos}-\ref{fig:lno_orb_pdos}) are calculated based on the converged wavefunctions. 

The structures for pristine \LCO and partially delithiated \LqCO are based on experimentally determined data from previous work~\cite{Hu2021_LCO_RIXS}. Constrained magnetization is used for \LqCO relaxation to get the magnitude of spin polarization on cobalt atoms that best fit the experiment measured spectrum. We note that the resulting magnitude of polarization and density of states agrees with the result calculated by Vienna ab-initio Simulation Package (VASP)~\cite{vasp1,vasp2}.

\begin{figure} 
	\centering
	\includegraphics[width=0.4\textwidth]{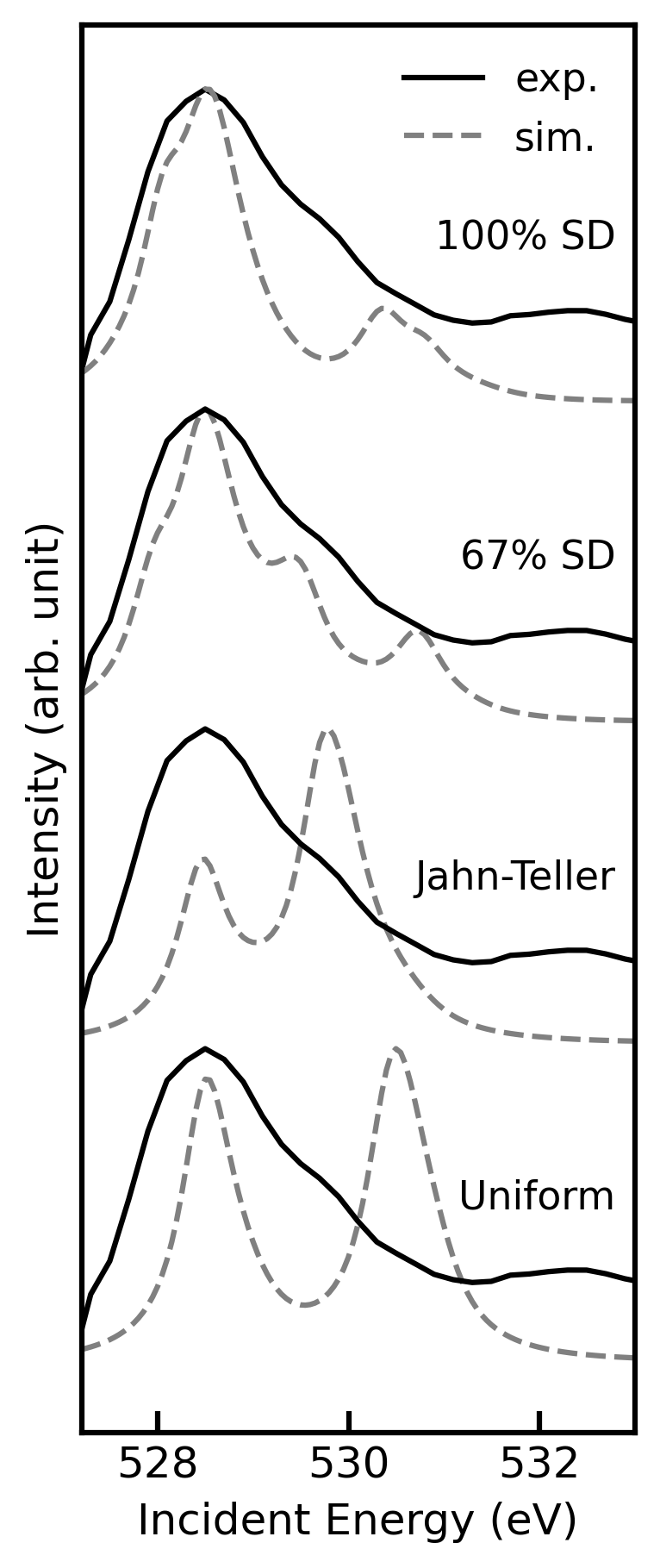} 

	\caption{\textbf{Calculated sXAS spectra different LiNiO\textsubscript{2} candidates.} {\sc ocean} simulations (dashed lines) of the O \Kedge sXAS for different \LNO candidate structures, compared to the experimental sXAS PFY measurements (solid lines). From bottom to top: Uniform, Jahn-Teller distorted, 67\% SD, and 100\% SD structures. All simulated spectra are normalized to the largest pre-edge peak. The simulated spectra of the 67\% SD structure shows the best qualitative agreement with the experimental measurement.}
	\label{fig:lno_xas_diff_struct} 
\end{figure}

Four different structures with distinct distortion modes were considered for pristine \LNO. A structure with (a) uniform \Nioct octahedra with $O3$ stacking ($\mathrm{R\bar{3}m}$ space group, denoted as Uniform) was identified through powder diffraction measured by Chien et al~\cite{Chien_LNO_struct}. We also considered structures with: (b) zigzag Jahn-Teller ($\mathrm{P2_1/c}$ space group, denoted Jahn-Teller) distortion, (c) an even distribution of Jahn-Teller and breathing modes ($\mathrm{C2}$ space group, denoted 67\% SD) and (d) a structure with only breathing mode ($\mathrm{P2/c}$ space group, denoted 100\% SD), based on Foyevtsova et al. theoretical calculations~\cite{Kateryna2019disp}. We selected 67\% SD structure for all the calculations in the main text. Structural analysis using EXAFS and O \Kedge XAS simulations are shown in Fig.~\ref{fig:lno_exafs} and Fig.~\ref{fig:lno_xas_diff_struct} , respectively, which confirm our choice of structure.

\begin{figure} 
	\centering
	\includegraphics[width=0.9\textwidth]{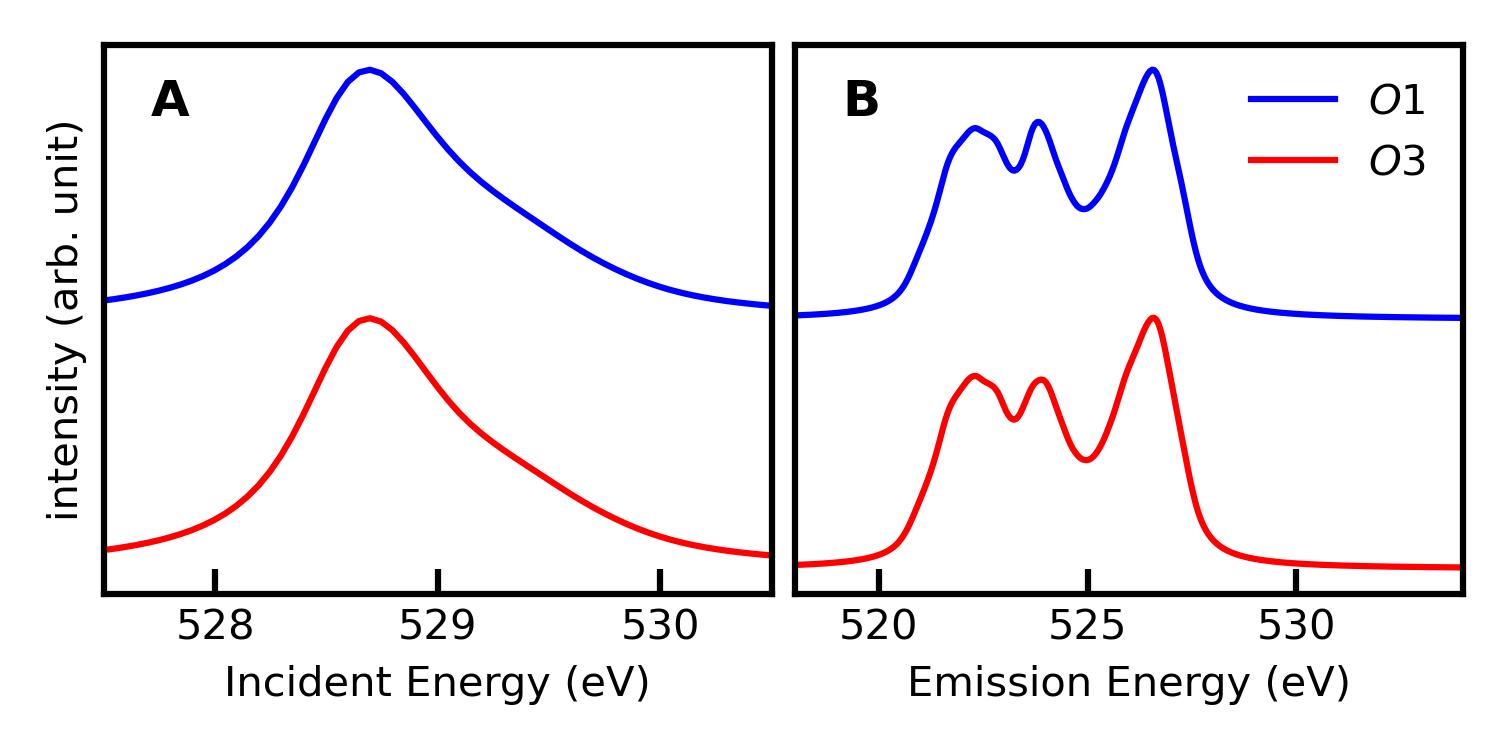} 

	\caption{\textbf{Calculated spectra for different NiO\textsubscript{2} stacking order.} {\sc ocean} simulation of $O1$ and $O3$ stacked \NO for O \Kedge (\textbf{A}) XAS (\textbf{B}) XES. The same global energy shift is used for both XAS and XES to fit experimental measurements. The resulting spectra from different stacking order are indistinguishable from each other and qualitatively similar to the experimental measurements, indicating that the electronic structure is insensitive to the stacking order.}
	\label{fig:lno_chg_calc} 
\end{figure}

The structure for charged \NO was based on experimental measurements from Chien et al~\cite{Chien_LNO_struct}. We considered both $O1$ and $O3$ stacking for the nickel oxide layers. The ground state energy of $O3$ stacked \NO is 6.2~meV lower then the $O1$ stacked \NO. This is consistent with experimental observation where $H3$ phase is a mixture of both $O1$ and $O3$ stacked structure~\cite{Chien_LNO_struct}. We note that the stacking structure has no effect on the calculated oxygen \Kedge XAS and XES (Fig.~\ref{fig:lno_chg_calc}).

\subsubsection*{EXAFS calculation} \label{sec:exafs}

\begin{figure} 
	\centering
	\includegraphics[width=0.8\textwidth]{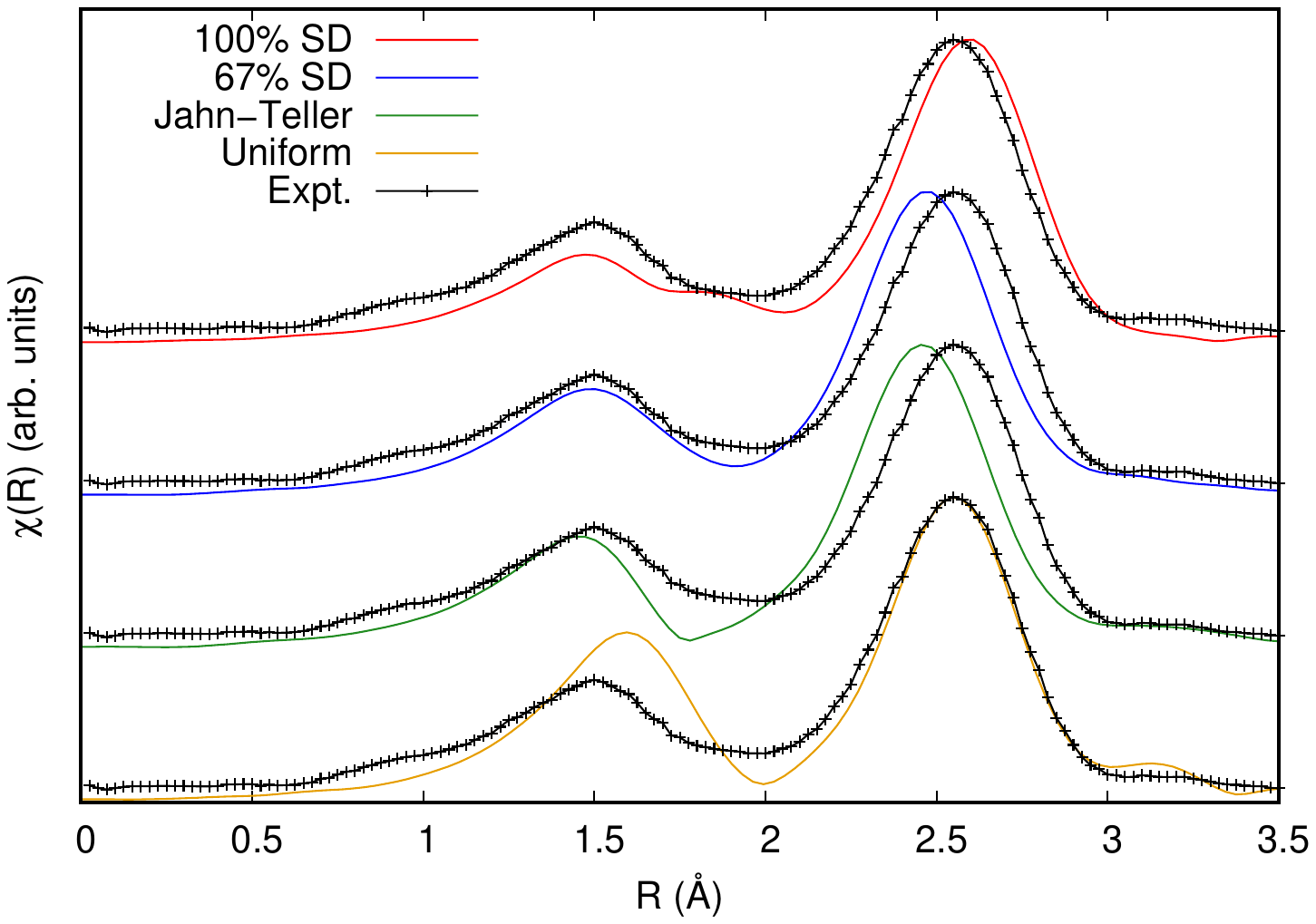} 

	\caption{\textbf{Fourier transformed simulated EXAFS oscillations $\boldsymbol{\chi}$(R) for the five candidate structures of LiNiO\textsubscript{2}.}}
	\label{fig:lno_exafs} 
\end{figure}

EXAFS (Fig.~\ref{fig:lno_exafs}) were calculated for all candidate \LNO structures using FEFF10~\cite{feff10}, and compared to experimental data digitally reproduced from ref.~\cite{Antonella_EXAFS}. The mean-square relative displacements (MSRDs) were calculated from the dynamical matrix following~\cite{vila_DMDW}. The dynamical matrix was calculated using VASP for the $\mathrm{R\bar{3}m}$ structure, and the extracted MSRDs were applied to similar paths for the EXAFS calculations of the other structures. The most relevant MSRD values were 0.0047~$\mathrm{\mathring{A}}^2$ and 0.00378~$\mathrm{\mathring{A}}^2$ for the Ni-O and Ni-Ni single scattering paths, respectively. These values are in good agreement with those extracted from experiment, for all of the structures except the $\mathrm{R\bar{3}m}$ structure, for which a much larger MSRD is required for the Ni-O paths to describe the first peak\cite{Antonella_EXAFS}. The results are similar enough to show that the 100\% SD, Jahn-Teller distorted, and 67\% SD structures are all consistent with experimental measurements given the errors in bond-length associated with DFT, as seen in previous work~\cite{Antonella_EXAFS}. Previous study has also shown that the 67\% SD structure agrees remarkably well with experimental \LNO neutron pair distribution function (nPDF) measurements as well~\cite{Kateryna2019disp}.

\subsubsection*{{\sc ocean} calculation} \label{ref:ocean}

The {\sc ocean} code (version 3.1.1)~\cite{OCEAN,OCEAN3} was used to calculate oxygen \Kedge XAS and RIXS for all structures. Input wavefunctions are generated using the QuantumEspresso DFT package using parameters consistent with the DFT ground state calculation. The convergence criterion for the electronic self-consistent field calculation was set to $10^{-10}$~Ry. Incoming and outgoing photon polarization along the x, y and z directions were considered. For O \Kedge, a 0.3 eV and 0.2 eV broadening is applied in the absorption and loss energy axis, respectively. A global energy shift for both XAS and RIXS was fitted to experiments.

\newpage

\begin{figure} 
	\centering
	\includegraphics[width=0.8\textwidth]{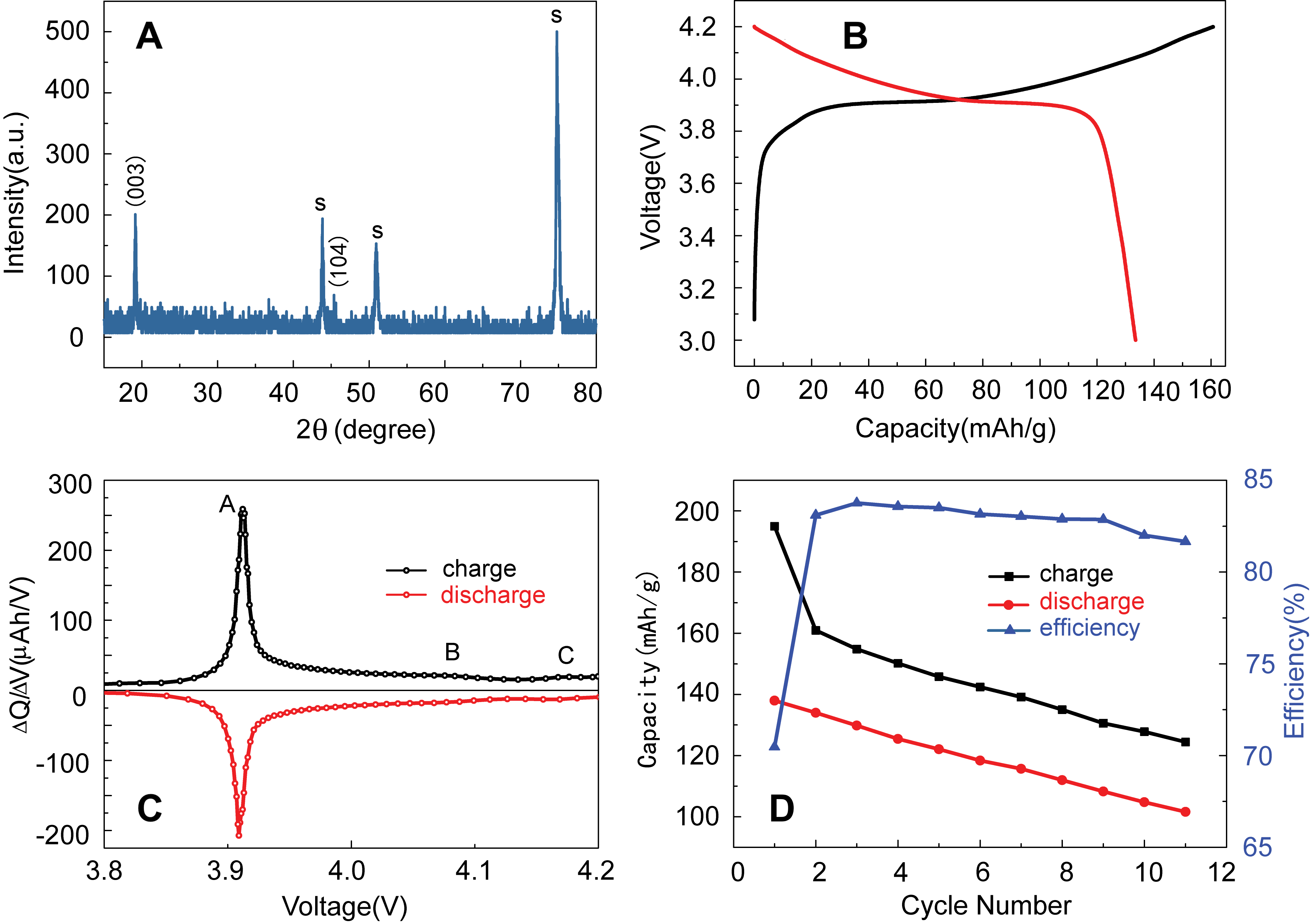} 

	\caption{\textbf{Characterization of \LxCO cell.} (\textbf{A}) XRD pattern of \LCO thin film on stainless steel substrate deposited at 500 centigrade oxygen pressure 10~Pa by PLD, (\textbf{B}) Typical charge-discharge curve of \LCO thin film, (\textbf{C}) Incremental capacity versus potential derived from charge-discharge curve, (\textbf{D}) Charge/discharge capacity and Coulombic efficiency versus cycle number of \LCO film electrochemically cycled in the voltage range between 3.0~V and 4.2~V.}
	\label{fig:lco_xrd} 
\end{figure}

\begin{figure} 
	\centering
	\includegraphics[width=0.8\textwidth]{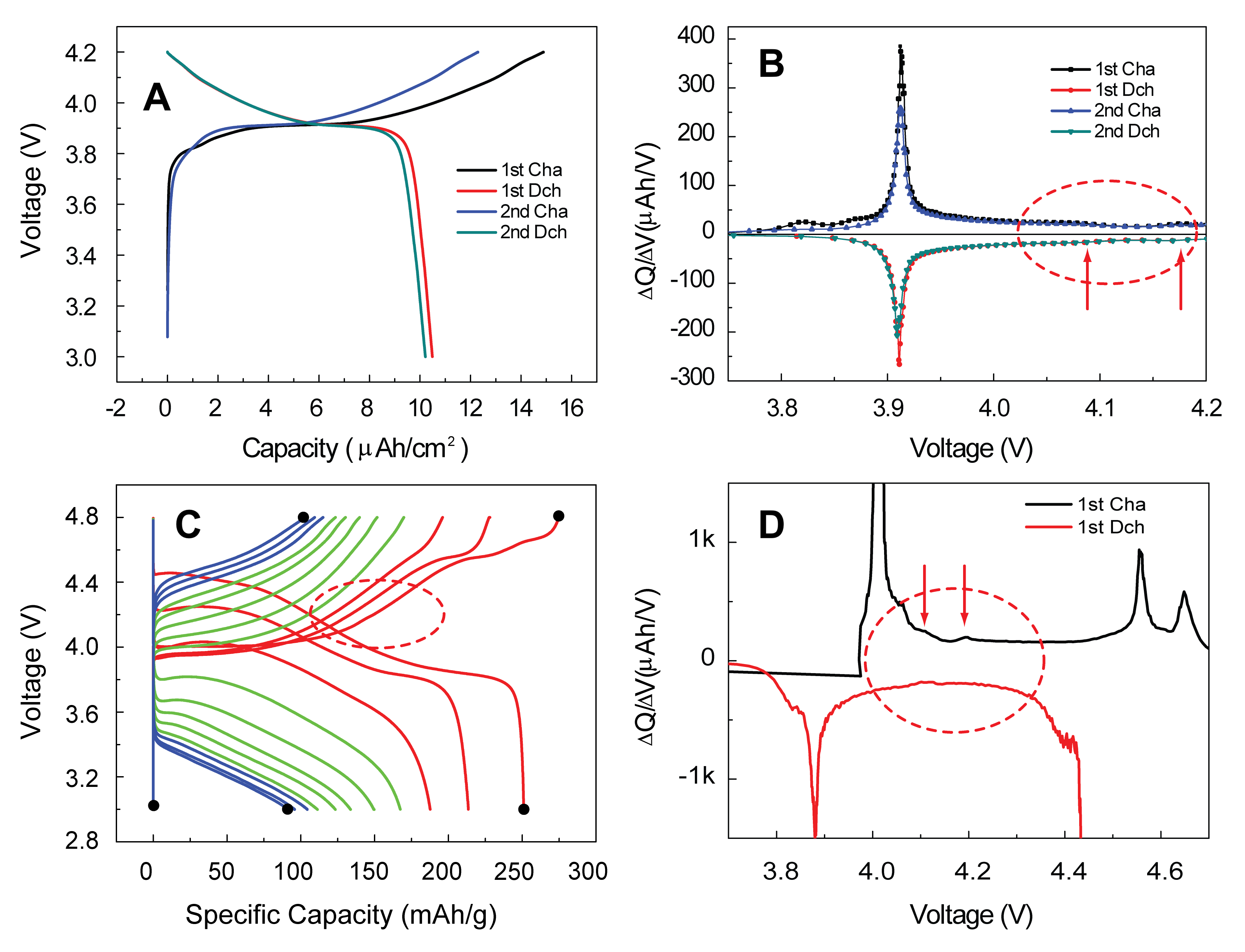} 

	\caption{\textbf{Electrochemical profiles of LiCoO\textsubscript{2} thin films.} (\textbf{A-B}) and standard electrodes from U.S. Department of Energy’s (DOE) CAMP (Cell Analysis, Modeling and Prototyping) Facility, Argonne National Laboratory (\textbf{C-D}). Circles and arrows indicate the voltage range where structural phase transformation takes place, leading to a redox kink that is emphasized in the dQ/dV profiles.}
	\label{fig:lco_elec_prof} 
\end{figure}

\begin{figure} 
	\centering
	\includegraphics[width=0.7\textwidth]{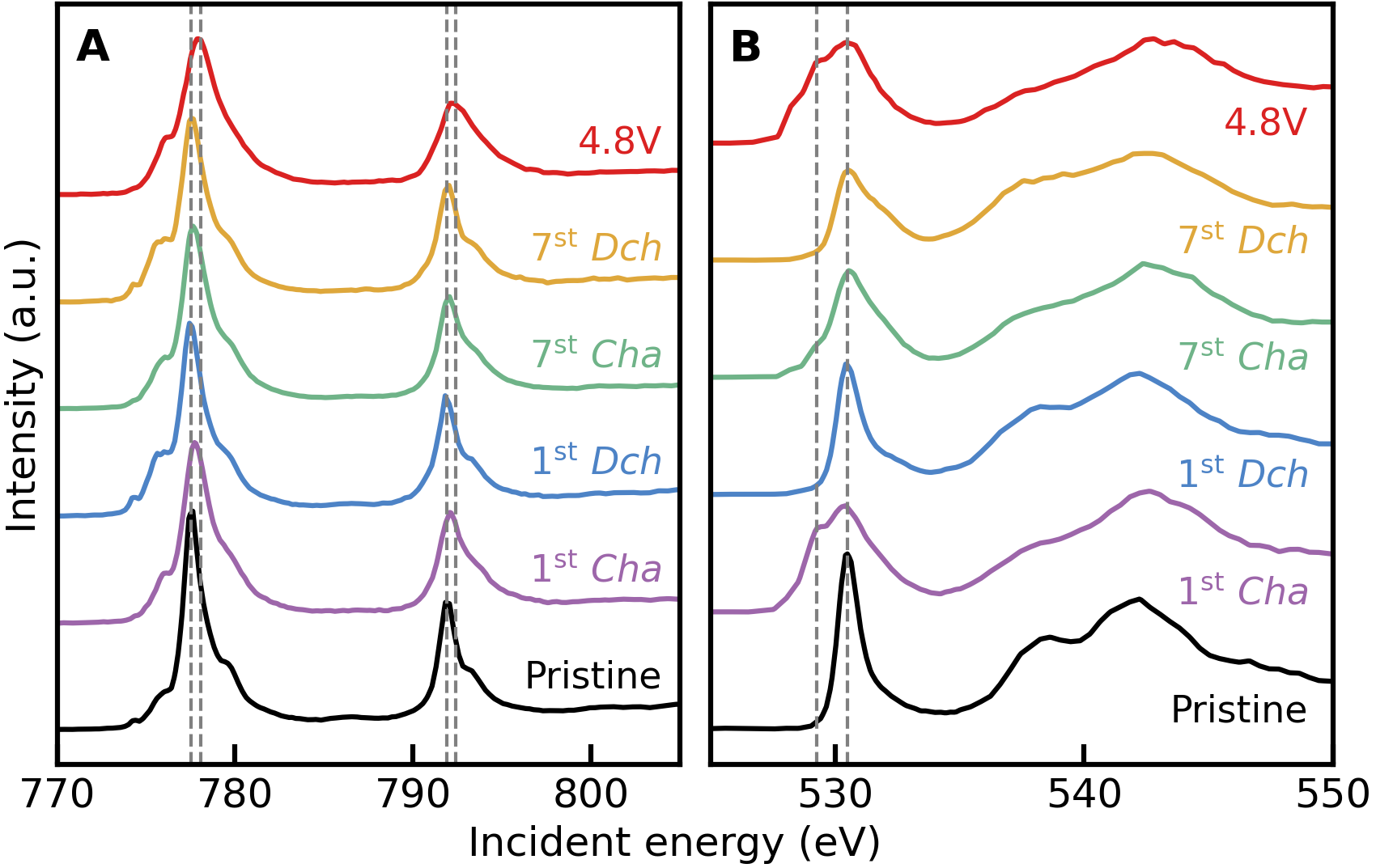} 

	\caption{\textbf{\textit{ex-situ} sXAS spectra.} (\textbf{A}) Co \Ledge sXAS and (\textbf{B}) O \Kedge sXAS collected in total electron yield (TEY) mode.}
	\label{fig:lco_xas_supp} 
\end{figure}

\begin{figure} 
	\centering
	\includegraphics[width=0.7\textwidth]{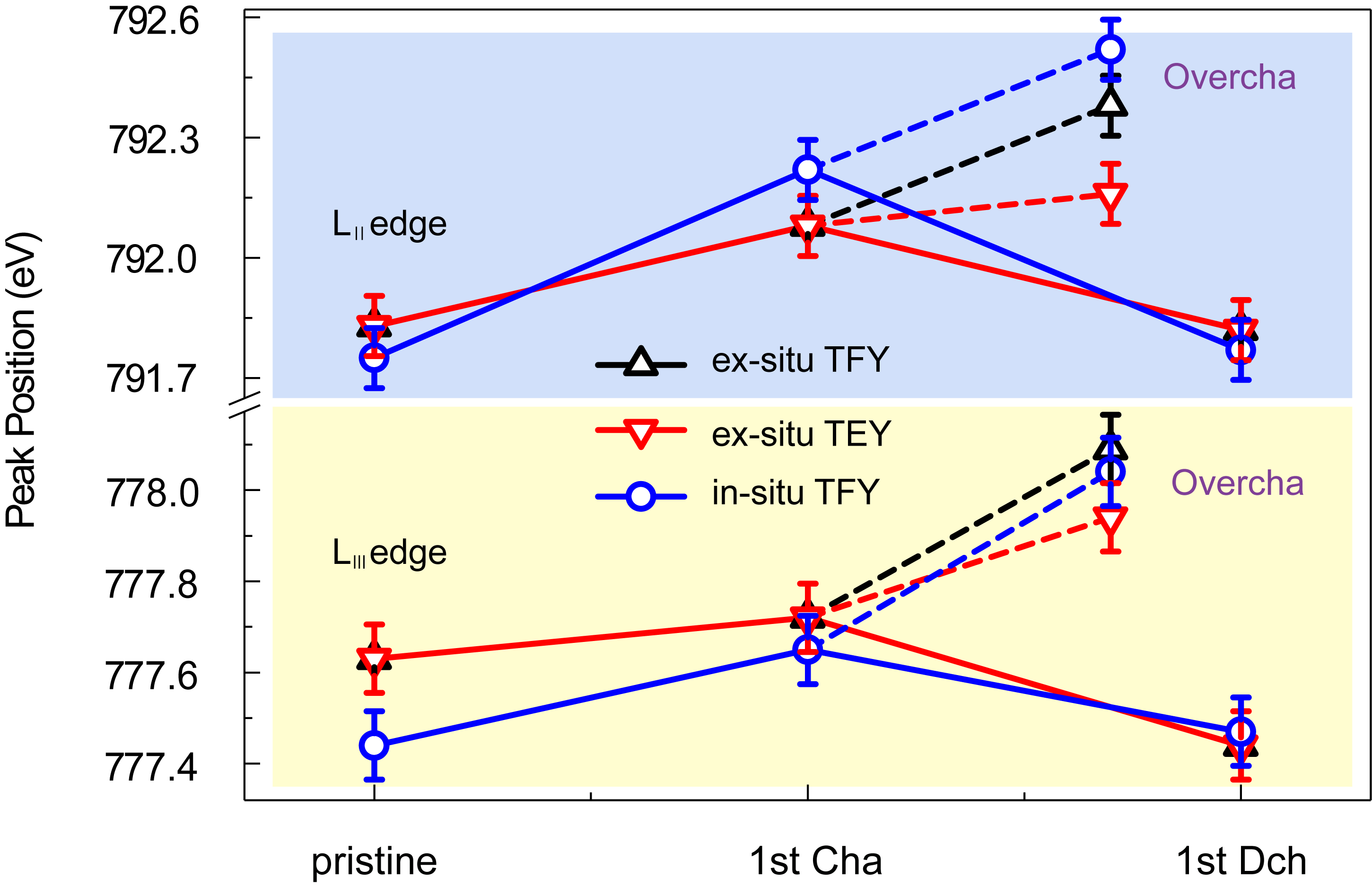} 

	\caption{\textbf{The energy shifts of the Co \textit{L\textsubscript{2}} and \textit{L\textsubscript{3}} white lines (main peaks) of LiCoO\textsubscript{2} electrodes upon electrochemical cycling.} “1\textsuperscript{st} Cha” refers to electrodes charged to the standard \LCO operation voltage of 4.2~V. “Overcha” are electrodes charged to 4.8~V. The \textit{in-situ}/\textit{operando} (Blue) and \textit{ex-situ} TFY (Black) spectra follow the same trend of peak shifts.}
	\label{fig:lco_edge_shift} 
\end{figure}

\begin{figure} 
	\centering
	\includegraphics[width=0.95\textwidth]{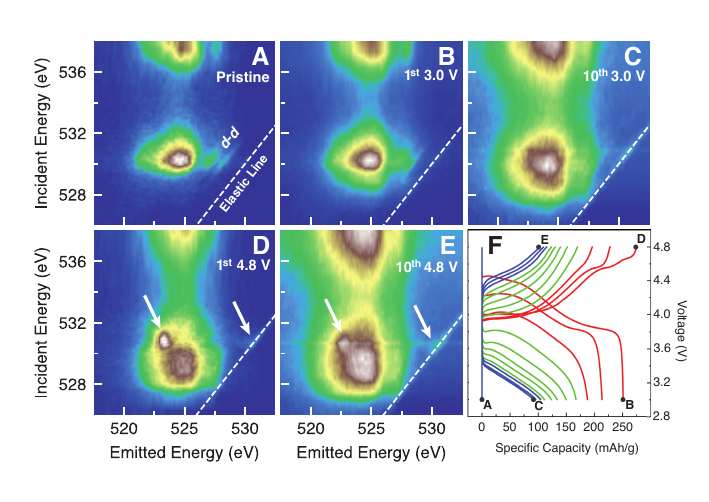} 

 	\caption{\textbf{O \textit{K-}edge RIXS of pristine, charged, and discharged LiCoO\textsubscript{2} after the 1st and 10th electrochemical cycles.} (\textbf{A-C}) are from \LCO electrodes in their discharged states. (\textbf{D,E}) are collected from \LCO electrodes charged to 4.8~V after the 1st and 10th cycles. (\textbf{F}) displays the electrochemical profile of the first 10 cycles and the sampling points.}
	\label{fig:lco_k_rixs_extend} 
\end{figure}

\begin{figure} 
	\centering
	\includegraphics[width=0.95\textwidth]{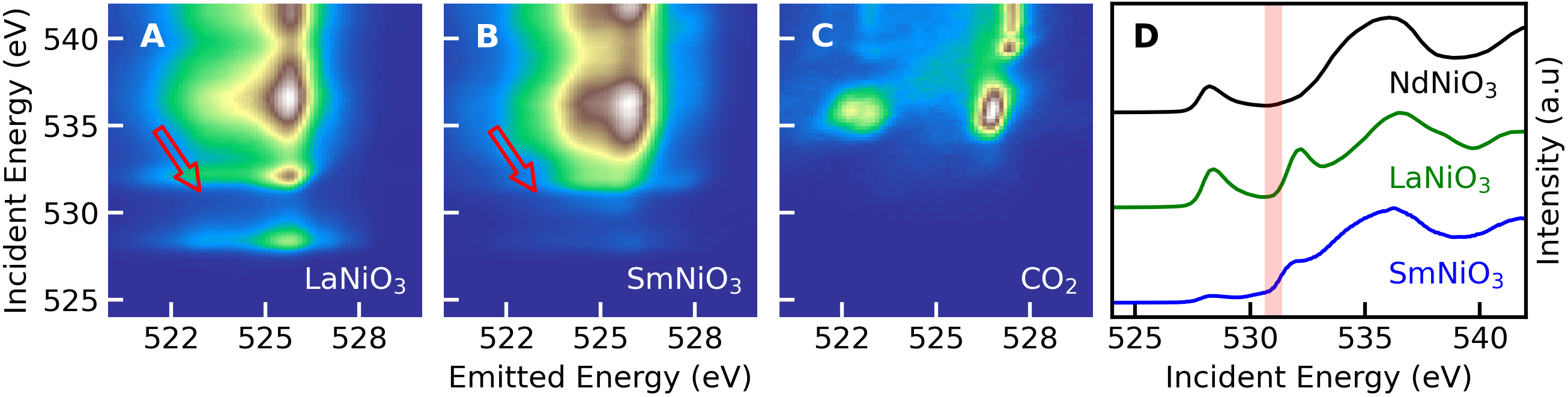} 
	\caption{\textbf{O \textit{K-}edge sXAS and RIXS of NCT systems.} (\textbf{A}) O \Kedge RIXS of typical NCT systems, LaNiO\textsubscript{3} and (\textbf{B}) SmNiO\textsubscript{3}. Hollow arrows indicate the energy range where one would expect features of oxidized oxygen species with O-O bonding formation. (\textbf{C}) O \Kedge RIXS of a highly covalent system, CO\textsubscript{2}. (\textbf{D}) O \Kedge sXAS of the same NCT systems collected in TFY mode. Shaded area indicate the energy range for features of oxidized oxygen species with O-O bonding formation.}
	\label{fig:ni_nct} 
\end{figure}

\begin{figure} 
	\centering
	\includegraphics[width=0.6\textwidth]{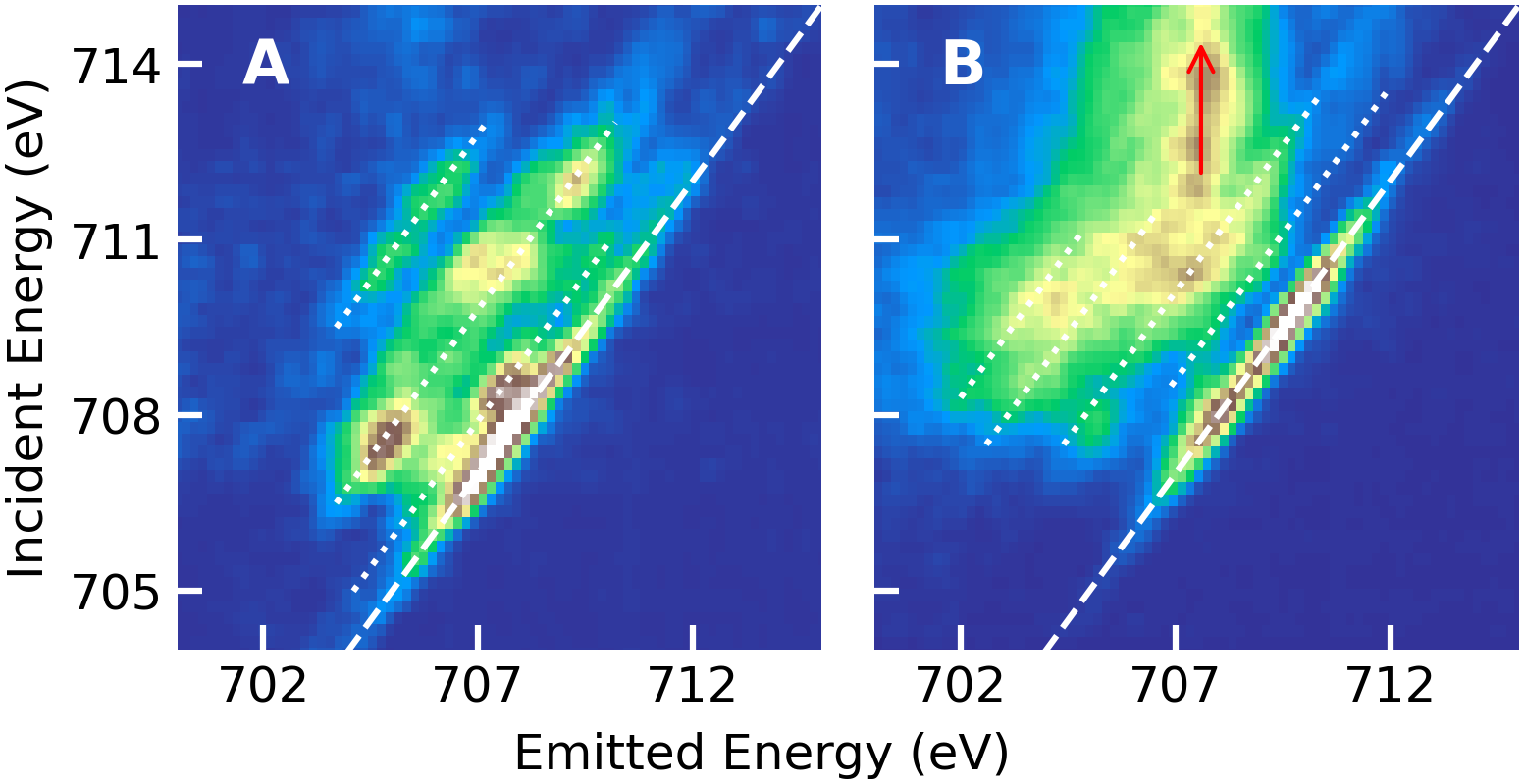} 

	\caption{\textbf{Fe \textit{L}\textsubscript{3}-edge RIXS.} (\textbf{A}) Pristine and (\textbf{B}) charged \LFPO electrodes. White dotted lines indicate the excitation features of localized Fe 3$d$ states, which could still be clearly identified at charged state although the RIXS feature of itinerant states. However, the fluorescence signals (indicated by red arrow) seen in charged \LFPO, is much enhanced.}
	\label{fig:lfpo_rixs} 
\end{figure}

\begin{figure} 
	\centering
	\includegraphics[width=0.9\textwidth]{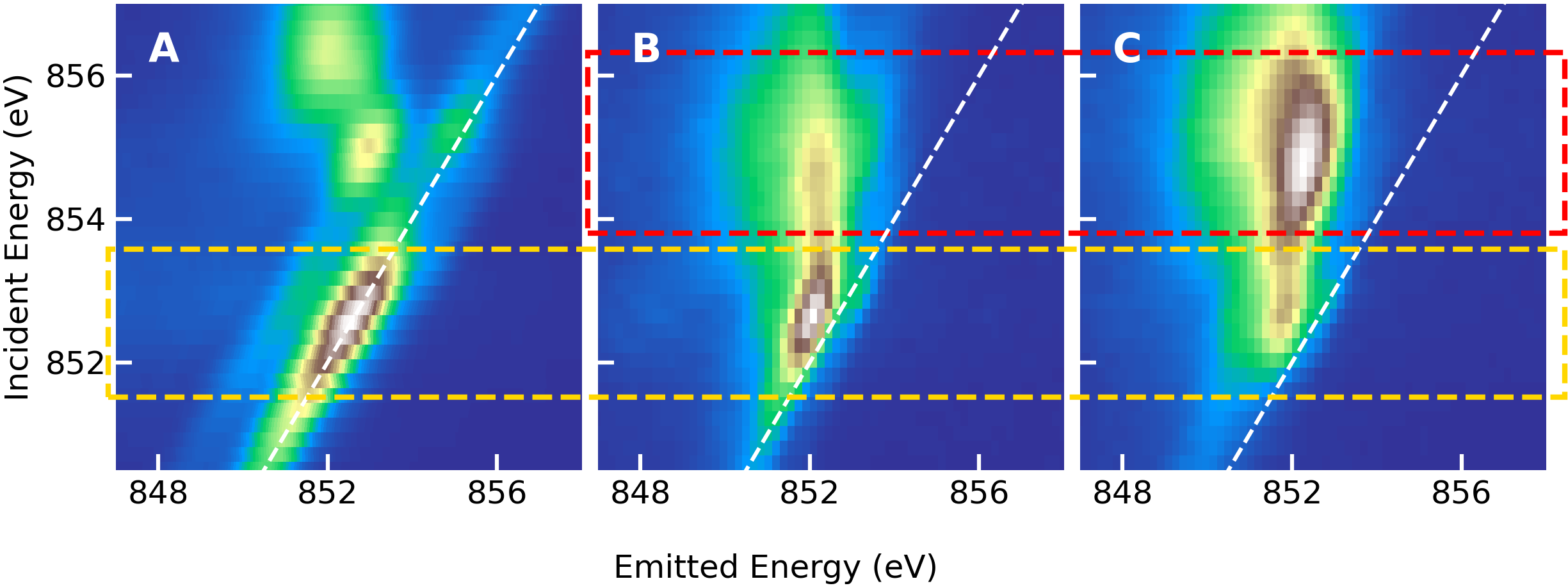} 

	\caption{\textbf{Ni \textit{L}\textsubscript{3}-edge RIXS for different Ni based compounds.} (\textbf{A}) NiO (\textbf{B}) pristine \LNO (also in Fig.~\ref{fig:lno_sim}A) (\textbf{C}) charged \LxNO (also in Fig.~\ref{fig:lno_sim}D). Rectangles are guides to the eye to indicate the two main parts of RIXS features of different ground state origins. The yellow rectangle highlights excitations associated with \Nioct that have long bonds and a PCT ground state. Excitations within the red rectangle are linked to NCT ground states, namely the Jahn-Teller distorted and the small \Nioct.}
	\label{fig:nio_rixs} 
\end{figure}

\newpage

\begin{table}[htbp] 
	\centering
	\caption{\textbf{Multiplet calculation parameters and orbital occupations for \LxCO.} Slater-Condon parameters ($F$ and $G$) were set by Hartree-Fock values~\cite{haverkort2005spinorbitaldegreesfreedom} scaled by 80\% for each configuration. Spin-orbit interactions (${\zeta}$) were set to their full HF value. $U_{dd}$, hybridizations ($t$), crystal field (${\Delta_O}$), and charge transfer energies (${\Delta}$) were determined by comparison to experiment. All units are in eV.}
    \renewcommand{\arraystretch}{1.1}
    \setlength{\tabcolsep}{2pt}
	\label{tab:lco_mult}
	\begin{tabular}{|c|ccccccccccccc|}
    \hline
    \multicolumn{14}{|c|}{\textbf{CTFHAM parameters for CoO\textsubscript{3} cluster}} \\\hline
    Config &$U_{dd}$ & $\Delta_O$ & $\Delta$ & $\zeta_{3d}$ & $F_{dd}^{2}$ & $F_{dd}^{4}$ & $\zeta_{2p}$ & $F_{pd}^{2}$ & $G_{pd}^{1}$ & $G_{pd}^{3}$ & $t_{t_{2g}}$ & $t_{e_g}$ & $t_{pp}$ \\\hline
     $2p^63d^7$ & 5.0 & 1.4 & 8.0 & 0.066 & 9.283 & 5.767 & - & - & - & - & 0.4 & 0.16 & 0.1 \\\hline
     $2p^53d^8$ & " & " & " & 0.083 & 9.916 & 6.166 & 9.748 & 5.807 & 4.315 & 2.526 & " & " & " \\\hline
     $2p^63d^6$ & 5.0 & 1.4 & 3.5 & 0.074 & 10.129 & 6.333 & - & - & - & - & 1.05 & 0.42 & 0.1   \\\hline
     $2p^53d^7$ & " & " & " & 0.092 & 10.737 & 6.715 & 9.746 & 6.319 & 4.758 & 2.707 & " & " & " \\\hline
     $2p^63d^5$ & 5.1 & 0.6 & -3.5 & 0.082 & 10.910 & 6.858 & - & - & - & - & 1.1 & 0.44 & 0.1   \\\hline
     $2p^53d^6$ & " & " & " & 0.101 & 11.498 & 7.277 & 9.746 & 6.835 & 5.220 & 2.973 & " & " & " \\\hline
     \multicolumn{14}{|c|}{\textbf{Ground state hole orbital occupations}} \\\hline
     Material & $d_{x^2-y^2}$ & $d_{z^2}$ & $d_{xy}$ & $d_{xz/yz}$ & $p_{xx,yy}$ & $p_{zz}$ & $p_{\pi}$  & \multicolumn{6}{|c|}{Contributions to $|\Psi_{GS}|^2$} \\\hline
     $\mathrm{CoO}$ & 0.95 & 0.95 & 0.36 & 0.36  & 0.008 & 0.006 & $5\times10^{-4}$ & \multicolumn{6}{|c|}{$0.97\langle d^7\rangle + 0.03 \langle d^8L\rangle$} \\\hline
     $\mathrm{LiCoO_2}$ & 1.61 & 1.64 & 0.02 & 0.02 & 0.25 & 0.20 & $2\times10^{-4}$ &  \multicolumn{6}{|c|}{$0.43 \langle d^6\rangle + 0.46 \langle d^7L\rangle + 0.11 \langle d^8L^2\rangle$} \\\hline
     $\mathrm{CoO_2}$ & 1.1 & 1.1 & 0.24 & 0.27 & 0.61 & 0.54 & 0.05  &  \multicolumn{6}{|c|}{ $0.22 \langle d^6L\rangle + 0.48 \langle d^7L^2\rangle + 0.24 \langle d^8L^3\rangle$} \\
		\hline
	\end{tabular}
\end{table}

\begin{table}[htbp] 
	\centering
	\caption{\textbf{Multiplet calculation parameters and orbital occupations for \LxNO.} Slater-Condon parameters ($F$ and $G$) were set by Hartree-Fock values~\cite{haverkort2005spinorbitaldegreesfreedom} scaled by 80\% for each configuration. Spin-orbit interactions (${\zeta}$) were set to their full HF value. $U_{dd}$, hybridizations ($t$), crystal field (${\Delta_O}$), and charge transfer energies (${\Delta}$) were determined by comparison to experiment. All units are in eV.}
    \renewcommand{\arraystretch}{1.}
    \setlength{\tabcolsep}{2pt}
	\label{tab:lno_mult}
	\begin{tabular}{|c|ccccccccccccc|}
    \hline
    \multicolumn{14}{|c|}{\textbf{CTFHAM parameters for NiO\textsubscript{3} cluster}} \\\hline
     Config &$U_{dd}$ & $\Delta_O$ & $\Delta$ & $\zeta_{3d}$ & $F_{dd}^{2}$ & $F_{dd}^{4}$ & $\zeta_{}$ & $F_{pd}^{2}$ & $G_{pd}^{1}$ & $G_{pd}^{3}$ & $t_{t_{2g}}$ & $t_{e_g}$ & $t_{pp}$ \\\hline
     $2p^63d^8$ & 5.5 & 1.0 & 5.0 & 0.083 & 9.786 & 6.078 & - & - & - & - & 0.6 & 0.3 & 0.1   \\\hline
     $2p^53d^7$ & " & " & " & 0.102 & 10.404 & 6.467 & 11.507 & 6.176 & 4.626 & 2.632 & " & " & " \\\hline
     $2p^63d^7$ & 5.5 & 1.0 & -1.5 & 0.091 & 10.621 & 6.635 & - & - & - & - & 1.0 & 0.5 & 0.1   \\\hline
     $2p^53d^6$ & " & " & " & 0.112 & 11.217 & 7.010 & 11.506 & 6.679 & 5.063 & 2.882 & " & " & " \\\hline
     $2p^63d^6$ & 5.6 & 1.5 & -5.0 & 0.101 & 11.395 & 7.155 & - & - & - & - & 1.15 & 0.575 & 0.6   \\\hline
     $2p^53d^5$ & " & " & " & 0.122 & 11.972 & 7.519 & 11.506 & 7.187 & 5.518 & 3.143 & " & " & " \\\hline
    \multicolumn{14}{|c|}{\textbf{Ground state hole orbital occupations}} \\\hline
     Material & $d_{x^2-y^2}$ & $d_{z^2}$ & $d_{xy}$ & $d_{xz/yz}$ & $p_{xx,yy}$ & $p_{zz}$  & $p_{\pi}$  & \multicolumn{6}{|c|}{Contributions to $|\Psi_{GS}|^2$} \\\hline
     $\mathrm{NiO}$ & 0.91 & 0.92 & 0.0 & 0.003 & 0.06 & 0.4 & 0.0  & \multicolumn{6}{|c|}{$0.85 \langle d^8\rangle + 0.15 \langle d^9L\rangle$} \\\hline
     $\mathrm{LiNiO_2}$ & 1.12 & 0.73 & 0.003 & 0.006 & 0.48 & 0.17 & 0.0 & \multicolumn{6}{|c|}{$0.16 \langle d^7\rangle + 0.57 \langle d^8L\rangle + 0.25 \langle d^9L^2\rangle$} \\\hline
     $\mathrm{NiO_2}$ & 1.03 & 1.03 & 0.005 & 0.005 & 0.65 & 0.62 & 0.0  & \multicolumn{6}{|c|}{$0.26 \langle d^7L\rangle + 0.50 \langle d^8L^2\rangle + 0.20 \langle d^9L^3\rangle$} \\\hline
	\end{tabular}
\end{table}

\begin{table}[htbp] 
	\centering
	\caption{\textbf{Average Bader charges (referenced to the neutral atomic species) for each element for the pristine and charged LiCoO\textsubscript{2}.} Co and O atoms have 2 different charge states in partially charged \LqCO given that some lithium atoms remains in the charged \LqCO computational cell. The atoms that are closer to the remaining Li atoms have smaller changes in the oxidation state. The largest charge contribution comes from oxygen atoms with 0.7 holes per lithium removed, whereas the rest of the added hole densities ($\sim$0.18) are distributed between Co and Li atoms.}
	\label{tab:lco_bader}
	\begin{tabular}{|c|c|c|c|}
    \hline
    \textbf{Element} & \textbf{LiCoO\textsubscript{2} charge} & \textbf{Li\textsubscript{1/3}CoO\textsubscript{2} charge} & \textbf{$\Delta$ Charge} \\ \hline
    Li & 0.902 & 0.904 & +0.002 \\ \hline
    Co & 1.373 & 1.428/1.529 & +0.055/+0.156 \\ \hline
    O & -1.137 & -1.139/-0.785 & -0.002/+0.352 \\ \hline
	\end{tabular}
\end{table}

\begin{table}[htbp] 
	\centering
	\caption{\textbf{Average Bader charges (referenced to the neutral atomic species) for each element for the pristine and charged LiNiO\textsubscript{2}.}  There is a 1\% difference in Bader charge for $O3$ and $O1$ stacking, thus we only report the Bader charge for the $O3$ stacked structure. Our Bader analysis shows that upon delithiation, pristine \LNO goes from a superposition of \deight+ \deightL+ \deightLt to charged \NO with a \deightLt ground state. Nickel atoms with longer Ni-O bonds gain additional holes, while those with shorter Ni-O bonds retain the same hole population upon charging.}
	\label{tab:lno_bader}
	\begin{tabular}{|c|c|c|c|}
    \hline
    \textbf{Element} & \textbf{LiNiO\textsubscript{2} charge} & \textbf{NiO\textsubscript{2} charge} & \textbf{$\Delta$ Charge} \\ \hline
    Li & 0.8955 & N/A & N/A \\ \hline
    Ni & 1.299/1.373/1.467 & 1.482 & +0.183/+0.109/+0.015 \\ \hline
    O & -1.138 & -0.741 & +0.397 \\ \hline
	\end{tabular}
\end{table}

\clearpage

\end{document}